\documentclass[longauth]{aa}  
\usepackage{xcolor}

\usepackage{graphicx}
\usepackage{txfonts}
\newcommand{\grb}{GRB\,190114C}

\begin{document}

   \title{GRB\,190114C in the nuclear region of an interacting galaxy\thanks{Partially based on \textit{Hubble Space Telescope} observations obtained under Director's Discretionary Time programme number 15684 (P.I.: Levan) and under programme number 15708 (P.I.: Levan). Partially based on Very Large Telescope observations obtained by the \textit{Stargate} Consortium under programme 0102.D-0662 (P.I.: Tanvir).
   Partially based on Atacama Large Millimeter  Array observations obtained under Director's Discretionary Time programme ADS/JAO.ALMA\#2018.A.00020.T (P.I.: de Ugarte Postigo).}}

   \subtitle{A detailed host analysis using ALMA, HST and VLT}

\author{
A. de Ugarte Postigo\inst{\ref{af:iaa},\ref{af:dark}}
\and
C.~C.~Th\"one\inst{\ref{af:iaa}}
\and
S.~Mart\'in\inst{\ref{af:eso},\ref{af:alma}}
\and
J.~Japelj \inst{\ref{af:apia}}
\and
A.~J. Levan \inst{\ref{af:imapp},\ref{af:warwick}}
\and
M.~J.~Micha{\l}owski\inst{\ref{af:pol}}
\and
J.~Selsing \inst{\ref{af:dawn},\ref{af:nbi}}
\and
D.~A.~Kann\inst{\ref{af:iaa}}
\and
S.~Schulze\inst{\ref{af:weiz}}
\and
J.~T.~Palmerio\inst{\ref{af:GEPI},\ref{af:IAP}}
\and
S.~D.~Vergani\inst{\ref{af:GEPI}}
\and
N.~R.~Tanvir \inst{\ref{af:leicester}}
\and
K.~Bensch\inst{\ref{af:iaa}} 
\and
S.~Covino\inst{\ref{af:Brera}}
\and
V.~D'Elia \inst{\ref{af:asi},\ref{af:inafoar}}
\and
M.~De Pasquale\inst{\ref{af:istanbul}}
\and
A.~S.~Fruchter\inst{\ref{af:stsci}}
\and
J.~P.~U.~Fynbo \inst{\ref{af:dawn},\ref{af:nbi}}
\and
D.~Hartmann \inst{\ref{af:dieter}}
\and
K.~E.~Heintz \inst{\ref{af:iceland}}
\and
A.~J.~van~der~Horst \inst{\ref{af:gwu},\ref{af:apsis}}
\and
L. Izzo \inst{\ref{af:iaa},\ref{af:dark}}
\and
P.~Jakobsson \inst{\ref{af:iceland}}
\and
K.~C.~Y.~Ng\inst{\ref{af:weiz}, \ref{af:grappa}}
\and
D.~A.~Perley \inst{\ref{af:ljmu}}
\and
A.~Rossi \inst{\ref{af:oas}}
\and
B.~Sbarufatti \inst{\ref{af:psu}}
\and
R.~Salvaterra \inst{\ref{af:iasf}}
\and
R.~S\'{a}nchez-Ram\'{\i}rez \inst{\ref{af:iaps}}
\and
D.~Watson \inst{\ref{af:dawn},\ref{af:nbi}}
\and
D.~Xu \inst{\ref{af:naoc}}
}

\institute{
Instituto de Astrof\'isica de Andaluc\'ia, Glorieta de la Astronom\'ia s/n, 18008, Granada, Spain\label{af:iaa}
\and
DARK, Niels Bohr Institute, University of Copenhagen, Lyngbyvej 2, DK-2100 Copenhagen \O, Denmark \label{af:dark}
\and 
 European Southern Observatory, Alonso de C\'ordova, 3107, Vitacura, Santiago 763-0355, Chile \label{af:eso}
 \and
Joint ALMA Observatory, Alonso de C\'ordova, 3107, Vitacura, Santiago 763-0355, Chile \label{af:alma}
 \and
Anton Pannekoek Institute for Astronomy, University of Amsterdam, Science Park 904, 1098 XH Amsterdam, The Netherlands\label{af:apia}
 \and
Department of Astrophysics/IMAPP, Radboud University Nijmegen, The Netherlands \label{af:imapp}
\and
Department of Physics, University of Warwick, Coventry CV4 7AL, UK \label{af:warwick}
\and
Astronomical Observatory Institute, Faculty of Physics, Adam Mickiewicz University, ul.~S{\l}oneczna 36, 60-286 Pozna{\'n}, Poland \label{af:pol}
\and
The Cosmic Dawn Center (DAWN) \label{af:dawn}
\and
Niels Bohr Institute, University of Copenhagen, Lyngbyvej 2, DK-2100 Copenhagen \O, Denmark \label{af:nbi}
\and 
Department of Particle Physics and Astrophysics, Weizmann Institute of Science, Rehovot 7610001, Israel \label{af:weiz}
\and
GEPI, Observatoire de Paris, PSL University, CNRS, 5 Place Jules Janssen, F-92190 Meudon, France\label{af:GEPI}
\and
Institut d’Astrophysique de Paris, Sorbonne Université, CNRS, UMR7095,  F-75014, Paris, France\label{af:IAP}
\and
Department of Physics and Astronomy, University of Leicester, LE1 7RH, UK \label{af:leicester}
\and
INAF - Osservatorio Astronomico di Brera, via E. Bianchi 46, I-23807 Merate (LC), Italy\label{af:Brera}
\and
ASI- Space Science Data Centre, Via del Politecnico snc, I-00133 Rome, Italy\label{af:asi}
\and
INAF - Osservatorio Astronomico di Roma, Via Frascati 33, I-00040 Monte Porzio Catone (RM), Italy\label{af:inafoar}
\and
Department of Astronomy and Space Sciences, Istanbul University, 34119, Beyaz\i t, Istanbul, Turkey\label{af:istanbul}
\and
Space Telescope Science Institute, 3700 San Martin Drive, Baltimore, MD21218, USA\label{af:stsci}
\and
Department of Physics and Astronomy, Clemson University, Clemson, SC 29634, USA \label{af:dieter}
\and
Centre for Astrophysics and Cosmology, Science Institute, University of Iceland, Dunhagi 5, 107, Reykjav\'ik, Iceland\label{af:iceland}
\and
Department of Physics, The George Washington University, 725 21st Street NW, Washington, DC 20052, USA\label{af:gwu}
\and
Astronomy, Physics, and Statistics Institute of Sciences (APSIS), The George Washington University, Washington, DC 20052, USA\label{af:apsis}
\and
GRAPPA Institute, University of Amsterdam, 1098 XH Amsterdam, The Netherlands\label{af:grappa}
\and
Astrophysics Research Institute, Liverpool John Moores University, 146 Brownlow Hill, Liverpool L3 5RF, United Kingdom\label{af:ljmu}
\and
INAF - Osservatorio di Astrofisica e Scienza dello Spazio, via Piero Gobetti 93/3, 40129 Bologna, Italy\label{af:oas}
\and
Department of Astronomy and Astrophysics, Pennsylvania State University, 525 Davey Laboratory, University Park, PA 16802, USA\label{af:psu}
\and
INAF - Istituto di Astrofisica Spaziale e Fisica Cosmica, via A. Corti 12, I-20133, Milano, Italy\label{af:iasf}
\and
INAF, Istituto di Astrofisica e Planetologia Spaziali, Via Fosso del Cavaliere 100, I-00133 Roma, Italy\label{af:iaps}
\and
CAS Key Laboratory of Space Astronomy and Technology, National Astronomical Observatories, Chinese Academy of Sciences, Beijing 100101, China\label{af:naoc}
}

   \date{Received September 10, 2019; accepted November 18, 2019}

 
  \abstract
   {GRB\,190114C is the first GRB for which the detection of very-high energy emission up to the TeV range has been reported. It is still unclear whether environmental properties might have contributed to the production of these very high-energy photons, or if it is solely related to the released GRB emission.
   }
   {The relatively low redshift of the GRB ($z=0.425$) allows us to study the host galaxy of this event in detail, and to potentially identify idiosyncrasies that could point to progenitor characteristics or environmental properties responsible for such a unique event.}
   {We use ultraviolet, optical, infrared and submillimetre imaging and spectroscopy obtained with HST, VLT and ALMA to obtain an extensive dataset on which the analysis of the host galaxy is based.}
   {The host system is composed of a close pair of interacting galaxies ($\Delta$v$=$50\,km\,s$^{-1}$), both of which are well-detected by ALMA in CO(3-2). The GRB occurred within the nuclear region ($\sim$170 pc from the centre) of the less massive but more star-forming galaxy of the pair. The host is more massive (log(M/M$_\odot$)$=$9.3) than average GRB hosts at that redshift and the location of the GRB is rather unique. The enhanced star-formation rate was probably triggered by tidal interactions between the two galaxies. Our ALMA observations indicate that both host galaxy and companion have a high molecular gas fraction, as has been observed before in interacting galaxy pairs.}
   {The location of the GRB within the core of an interacting galaxy with an extinguished line-of-sight is indicative of a denser environment than typically observed for GRBs and could have been crucial for the generation of the very-high-energy photons that were observed.}


   \maketitle
%

\section{Introduction}

Gamma-ray burst (GRB) host-galaxy studies have been important to constrain the properties and nature of the progenitor objects \citep{Hjorth2012,Vergani2015,Kruehler2015,Perley2016A,Perley2016B,Palmerio2019}.
Different types of progenitors are expected to be found in different types of galaxies, or in particular regions within them \citep{Fruchter06,Kelly2008,Svensson2010,Modjaz2011,Sanders2012,Kelly2014,Lyman2017,Japelj2018}. Most of the studies performed on GRB hosts have been based on observations in the optical and near-infrared range, where they have traditionally been more feasible. Until recently, observations of the host galaxies of GRBs at longer wavelengths were limited to small samples that were necessarily biased towards the very brightest objects due to the sensitivity of the observatories \citep{Tanvir2004,Hatsukade2011,Michalowski2012,deugarte2012a}, and by the limited availability of space observatories \citep{Hunt2014,Schady2014}. In particular, one of the most relevant ranges of study lies at millimetre and submillimetre wavelengths, where emission of cold dust within galaxies can be observed.
This wavelength range also covers many of the interesting atomic and molecular features associated with the clouds where GRB progenitors are expected to form. Furthermore, light emitted in this wavelength range is not affected by dust obscuration, enabling us to penetrate the densest regions of the galaxies. In the last years, new powerful millimetre and submillimetre observations, such as those performed by ALMA or NOEMA, are allowing us to study in detail GRBs and their galaxies using photometric \citep{Wang2012,Hatsukade2014,SanchezRamirez2017,Laskar2018}, spectroscopic \citep{Michalowski2016,deugarte2018,Hatsukade2019} and polarimetric \citep{Laskar2019} techniques.

  \begin{figure*}[!ht]
   \centering
   \includegraphics[width=\textwidth]{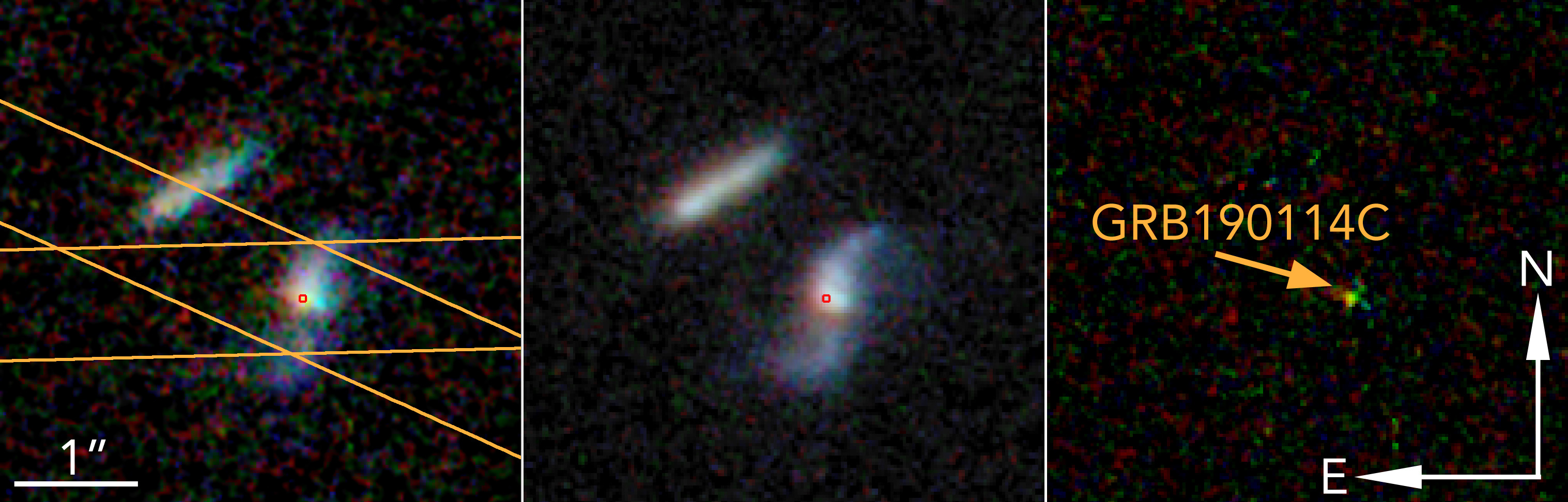}
   \caption{Colour composites obtained from the combination of the $F606W$, $F775W$, and $F850LP$ HST observing bands. The left image is made from the 11 February data, and the central one from the 28 August. The right panel is the subtraction of both frames, showing that the optical counterpart was still well detected in February. The location of the GRB is indicated with a red circle. The orange lines in the left panel indicate the position of the slits during the two X-shooter observations.}
         \label{Fig:fc}
   \end{figure*}

GRB\,190114C, localised \citep{Gropp2019} by the \textit{Neil Gehrels Swift Observatory} (\textit{Swift} hereafter; \citealt{Gehrels2004}), is the first GRB with confirmed very-high energy (VHE) emission (up to one TeV), and is thus a milestone in high-energy astrophysics \citep{Mirzoyan2019}. The GRB had one of the highest peak photon fluxes detected by BAT \citep{Krimm2019}, and was found to be at a redshift of $z=0.425$ \citep{Selsing2019,Castro-Tirado2019}. The sightline was significantly extinguished \citep{Kann2019}. Despite this, the early afterglow was exceedingly bright \citep{Lipunov2019}. Archival imaging from Pan-STARRS revealed what was later confirmed to be an underlying host galaxy \citep{deUgarte2019}.

Although HE $\gamma$-rays, as have been observed by \textit{Fermi}-LAT, can be caused by the high-energy tail of bright afterglow emission (see \citealt{Ajello2019} for an overview of observations and their interpretation), no VHE photons have been observed exceeding $\approx160$ GeV (rest-frame; see \citealt{Takahashi2019ICRC} for some very recent results). It has been proposed that the production of VHE emission could be due to other processes involved in the interaction of the burst ejecta with the circumburst medium~(see e.g. \citealt{Bottcher1998,Peer2005,Fan2008,Wang2010}; for reviews:~\citealt{Inoue2013,Nava2018}).
If a particularly dense environment was related to the VHE emission, one would expect to find VHE-emitting events in a particular region of the galaxy, where such conditions would be more favourable, such as the core of the galaxy or locations with a large amount of line-of-sight absorption\footnote{The recently discovered, very nearby GRB 190829A \citep{Fermi2019GCN,Dichiara2019GCN}, which has also been detected at extremely high energies \citep{deNaurois2019GCN}, also lies behind a highly extinguished line-of-sight (\citealt{Heintz2019GCN}, J. Bolmer, priv. comm.).}.  

In this paper we present optical, near-infrared, and submillimetre photometry and spectroscopy of the host galaxy of GRB\,190114C. We explore the peculiarities of this event and any distinctive factor that could have favoured the generation of the VHE photons that were observed.

Sect.~2 presents the observational data, Sect.~3 the results of our imaging and spectroscopic analyses. In Sect.~4 we discuss the implications of our findings. Sect.~5 presents our conclusions. Throughout the paper we use a cosmology ${H_0} = 71$\,km\,s$^{-1}$\,Mpc$^{-1}$, ${\Omega_{\rm M} = 0.27}$, and $\Omega_{\Lambda} = 0.73$ \citep{Spergel2003}. Unless specifically stated, the errors provided within this paper are 1-$\sigma$.


\section{Observations}

\subsection{Optical and near-infrared imaging from HST with ACS and WFPC3}

The {\it Hubble Space Telescope} (\textit{HST}) observed the field of GRB\,190114C using the Advanced Camera for Surveys \citep[ACS][]{Sirianni2005} at three epochs, within the context of the Director's Discretionary Time (DDT) programme number 15684 (P.I.: Levan), searching for the optical counterpart. The field was observed in four optical bands: $F475W$, $F606W$, $F775W$, and $F850LP$. Observations in all bands were obtained on 11 February 2019 and in $F475W$ and $F850LP$ also on 12 March 2019 
(the additional bands were not obtained at this epoch due to a guide-star acquisition failure). A visit with the missing filters $F606W$ and $F775W$ was obtained on 28 June 2019. 
To complete the dataset, a final visit was obtained on 28 August 2019 through the regular time programme number 15708 (P.I.: Levan). We obtained imaging in $F606W$, $F775W$, and $F850LP$ optical filters with ACS, and $F110W$, $F160W$ infrared filters with the Wide Field Camera 3 \citep[WFC3,][]{MacKenty2010}.
Data were first corrected for charge transfer efficiency (CTE) corrections to remove significant CTE features due to the harsh radiation environment, as well as the removal of bias striping. Processed images were then reduced and combined via {\tt astrodrizzle} with a final pixel scale of 0\farcs025/pixel. Images were co-aligned by the use of point sources in common to the different filters. 

The GRB is located near the core of its host galaxy, at $0\farcs030\pm0\farcs010$ (or a projected distance of $170\pm60$ pc) towards the Southeast of the galaxy's light centre. This is shown by image subtraction in the $F775W$ and $F850LP$ bands, where both afterglow and host are well-detected, with consistent offsets in both bands.

The host galaxy has a projected size of $1\farcs6\times0\farcs8$, equivalent to $9.1\times4.5$~kpc, as measured from the 3-$\sigma$ detection contours of an image combining all the visible bands. We also measure a radius of the galaxy of $r_{50}=0\farcs34=1.90$ kpc (containing 50\% of the light), or $r_{80}=0\farcs62=3.45$ kpc (containing 80\% of the light). These radii have been determined using the method described by \citet{Blanchard2016}.

In addition to the host galaxy of the GRB an additional galaxy is detected at a distance of $1\farcs2\pm0\farcs1$ (see Fig.~\ref{Fig:fc}). In this paper we confirm that both galaxies lie at the same redshift thanks to the ALMA spectroscopy described in Sect.~\ref{sect:alma}. At a redshift of $z=0.425$ this is equivalent to a projected distance of $6.8\pm0.6$ kpc. 
 
The counterpart is detected in the reddest bands during the first epoch (Acciari et al. 2019, in press), being particularly prominent in the $F775W$ band. In this paper we use the data from the fourth epoch, in which the GRB contribution is negligible, to study the host galaxy. The photometry was performed with SExtractor \citep{Bertin1996} using elliptical apertures with sizes based on the Kron radius \citep{Kron1980}, and using the zero points provided by the Space Telescope Science Institute for each of the filters at the specific epoch of the observation \citep{Sirianni2005}. The values obtained using this method (see Table~\ref{Tab:hostSED}) were compared to other aperture photometry methods, yielding similar results.

\subsection{Near-infrared imaging from VLT/HAWK-I}
More than two months after the GRB, on 23 March 2019, the field was observed in the near-infrared (NIR) bands with HAWK-I \citep[High-Acuity Wide-field K-band Imager;][]{pirard2004} 
mounted on the fourth Unit Telescope (UT4) of the Very Large Telescope (VLT), managed by the European Southern Observatory (ESO). The observation was performed within the {\it Stargate} collaboration programme 0102.D-0662 (P.I.: Tanvir). Both galaxies are well-detected and resolved in the three observing bands $J$, $H$, and $K_S$, with a seeing of 0\farcs60, 0\farcs55, and 0\farcs60, respectively. This allowed us to perform independent photometry of both the host and the companion galaxy. Data reduction was performed with a self-made pipeline based on IRAF routines. We obtained seeing-matched aperture photometry calibrated with respect to field stars from the 2MASS catalogue \citep{Skrutskie2006AJ}. This photometry is shown in Table~\ref{Tab:hostSED}. 

\subsection{Ultraviolet and infrared archival photometry from \textit{GALEX} and WISE}
We complement the host-galaxy photometry with catalogue ultraviolet observations from \textit{GALEX} \citep{Bianchi2011} and infrared observations from \textit{WISE} \citep{Wright2010}. Unfortunately, the spatial resolutions of \textit{GALEX} (4\farcs2 in the FUV and 5\farcs3 in the NUV band) and of \textit{WISE} (6\farcs1 in the 3.4 $\mu$m and 6\farcs4 in the 4.6 $\mu$m band) are not enough to resolve the two objects, and we can only obtain combined photometry of both galaxies.

The \textit{GALEX} photometry was obtained from the \textit{GALEX} Merged Catalog (MCAT) GR6/7 Data Release\footnote{http://galex.stsci.edu/GR6/}.
For \textit{WISE} observations we used the photometry from the AllWISE source catalogue\footnote{https://irsa.ipac.caltech.edu/Missions/wise.html}, transforming them from Vega to AB system with the corresponding AB-to-Vega offsets of 2.699 and 3.339 mag in the 3.4 and 4.6 $\mu$m bands, respectively \citep{Tokunaga2005}.

\begin{table}
\caption{Magnitudes (AB system) for the host galaxy, not corrected for Galactic extinction. T-T$_0$ indicates the epoch of the observation in days since the burst onset, where {\it pre} indicates data obtained in a pre-explosion epoch.
}
\label{Tab:hostSED}
\centering        
\small             
\begin{tabular}{l | c | c | c | c}       
\hline\hline                 
Filter	& Source	    & T-T$_0$0  & Host Mag          & Comp. Mag     \\
        & 	            & (day)  &                   &                \\
\hline                       
FUV     & {\it GALEX}   & pre & \multicolumn{2}{c}{24.60$\pm$0.32}    \\
NUV     & {\it GALEX}   & pre & \multicolumn{2}{c}{24.12$\pm$0.15}    \\ 
\hline                       
F475W   & {\it HST}/ACS  & 221 & 23.10$\pm$0.28    & 24.34$\pm$0.31    \\
F606W   & {\it HST}/ACS  & 221 & 22.31$\pm$0.16    & 22.95$\pm$0.21    \\
F775W   & {\it HST}/ACS  & 221 & 21.88$\pm$0.19    & 22.21$\pm$0.22    \\
F850LP  & {\it HST}/ACS  & 221 & 21.60$\pm$0.24    & 21.86$\pm$0.27    \\
F110W   & {\it HST}/WFC3 & 221 & 21.24$\pm$0.10    & 21.46$\pm$0.11    \\
{\it J} & VLT/HAWK-I     & 63 & 21.27$\pm$0.07    & 21.18$\pm$0.07    \\
F160W   & {\it HST}/WFC3 & 221 & 21.10$\pm$0.10    & 21.26$\pm$0.11    \\
{\it H} & VLT/HAWK-I     & 63 & 20.83$\pm$0.11    & 20.95$\pm$0.10    \\
{\it K$_S$}& VLT/HAWK-I  & 63 & 20.69$\pm$0.20    & 20.61$\pm$0.20    \\ 
\hline                       
3.4 $\mu$m & {\it WISE}  & pre & \multicolumn{2}{c}{20.11$\pm$0.11}   \\
4.6 $\mu$m & {\it WISE} & pre & \multicolumn{2}{c}{20.65$\pm$0.40}    \\
\hline                                   
\end{tabular}
\end{table}

\subsection{VLT/X-shooter spectroscopy}

Spectroscopy of GRB\,190114C was obtained with the three-arm X-shooter echelle spectrograph \citep{Vernet2011}, mounted on UT2 of VLT (programme 0102.D-0662, P.I.: Tanvir). Two epochs of X-shooter observations were obtained, beginning respectively 4.8 hr and 76 hr after the {\it Swift} trigger (see Th\"one et al. in prep. for more details). The initial results were reported by \cite{Kann2019}. The first observation was carried out using the 1\farcs0/0\farcs9/0\farcs9 slit, with observations lasting $4\times600$ s. The second observation consisted of an $8\times1200$ s spectroscopic integration with the 1\farcs0/0\farcs9/0\farcs9JH slit. The observations were performed using a nodding scheme with a 5\farcs0 nodding throw. Because the host is small and the surrounding regions are mostly empty, this allows for a very clean background subtraction. In this paper we use the second observation to perform emission-line analysis of the host galaxy of GRB\,190114C. This second epoch was divided into two executions of a $4\times1200$ s observing block. The slit was aligned with the parallactic angle at the beginning of each execution (see Fig.~\ref{Fig:fc}), and all the data were then combined to produce the spectrum that we used in this paper. The combined observation included the core of the host galaxy in the slit, with no significant contribution coming from the companion galaxy. The resulting spectrum covers the spectral range between 3200 and 18000 {\AA}.

  \begin{figure}[!ht]
   \centering
   \includegraphics[width=7cm]{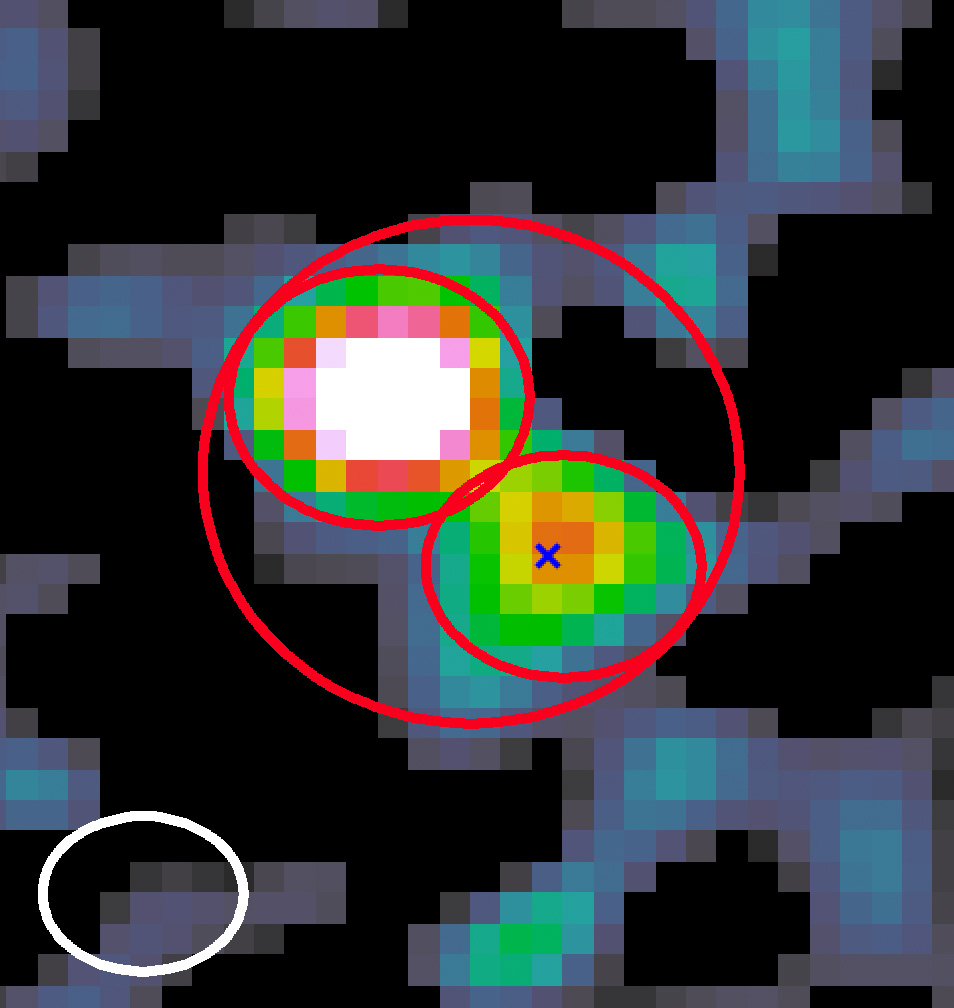}
   \caption{Image of the CO(3-2) emission obtained by ALMA. The spatial resolution (indicated by the white ellipse at the lower left) is just enough to resolve the two interacting galaxies. The red elliptical regions indicate the apertures used to measure the fluxes shown in Table~\ref{table:co}. The blue cross indicates the location of the afterglow.
 }
         \label{Fig:alma}
   \end{figure}

\begin{figure*}
	\centering
    \includegraphics[width=\textwidth]{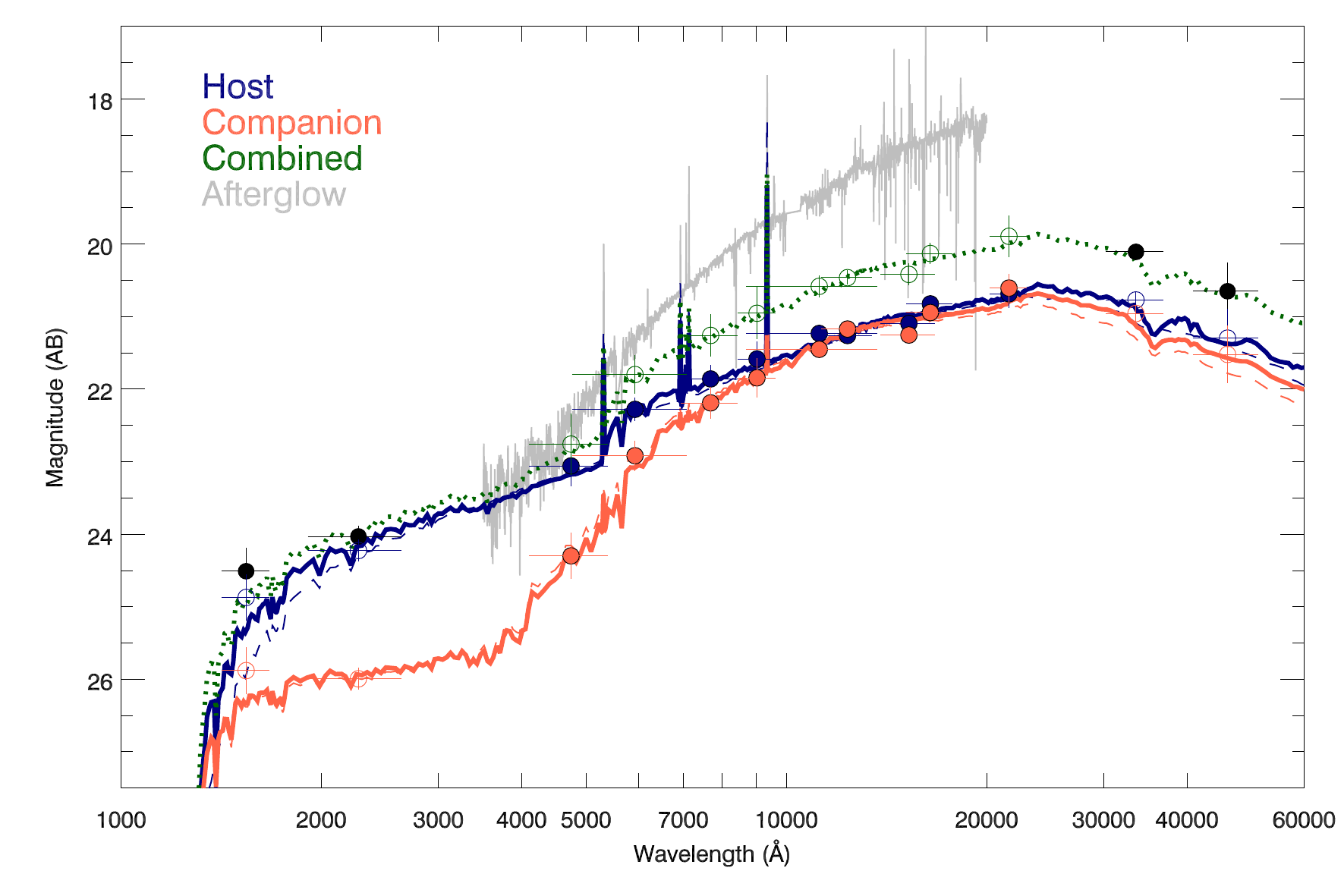}
	\caption{SED of the host galaxy of GRB\,190114C (blue) and its companion (red). The dashed lines indicate the fits without considering the \textit{GALEX} and \textit{WISE} photometry, where the two objects are blended. The solid lines are the fits considering also that photometry. In green we show the combined light of both galaxies. The spectrum of the second visit of X-shooter, which is afterglow-dominated above 5000 {\AA}, is plotted in grey.}
	\label{fig:hostsed}
\end{figure*}

\subsection{ALMA CO(3-2) imaging and spectroscopy}
\label{sect:alma}

Millimetre observations, tuned to cover the CO(3-2) line at the redshift of the GRB (i.e. 242.663~GHz), were carried out with the Atacama Large Millimeter Array (ALMA) observatory in Band 6, within the context of DDT programme ADS/JAO.ALMA\#2018.A.00020.T (P.I.: de Ugarte Postigo). Five individual executions were performed in three independent epochs ranging between January 17 and 18, 2019. The configuration used 47-48 antennas with baselines ranging from 15~m to 313~m ($12-253~k\lambda$ at the observed frequency). Each observation consisted of 43 min integration time on source with average weather conditions of precipitable water vapour $\sim3-4$~mm. The correlator was configured for a central observed frequency of 235.0487~GHz. The data were calibrated within CASA (Common Astronomy Software Applications, version 5.4.0, \citealt{McMullin2007}) using the pipeline calibration. The spatial resolution of the spectral data cube obtained by the pipeline products that combined all five executions was $1\farcs16\times0\farcs867$ (Position Angle $-87.9^\circ$), which was just enough to resolve both host galaxy and companion and perform independent analysis. The spectroscopic measurements were also analysed within CASA. The flux calibration was performed using the quasar J0522-3627 in the first two epochs and with J0423-0120 for the last one. 

The CO(3-2) zero-moment image is shown in Fig.~\ref{Fig:alma}.
To image the CO emission, continuum subtraction was performed in the UV visibilities using line free channels on both sides of CO(3-2). The spectral ranges used to fit the continuum were $241.839-242.433$ and $242.800-243.574$~GHz, which corresponds to a line emitting region of $\sim$450~km~s$^{-1}$. In order to ensure that the CO map in Fig.~\ref{Fig:alma} was not contaminated by residual continuum, we integrated several random velocity ranges within the considered line free regions. No significant residual continuum was detected on those verification maps.

\section{Results}

\subsection{SED fits}

In Table~\ref{Tab:hostSED} we show the observed magnitudes of the host galaxy of GRB\,190114C and its companion, based on the datasets described in Section 2. These observations were corrected for Galactic extinction ($\textnormal{E}(B-V)=0.011$, \citealt{Schlafly2011}) before proceeding with the SED fitting.

We used \texttt{LePhare} \citep{Arnouts1999,Ilbert2006} to fit these photometric data with a set of galaxy templates based on the models of \citet{Bruzual2003} and a Calzetti extinction law \citep{Calzetti2000}. \texttt{LePhare} uses an exponentially declining star-formation history (SFH) and a Chabrier initial mass function \citep[IMF,][]{Chabrier2003}.

To be able to include the unresolved photometry provided by \textit{GALEX} and \textit{WISE}, we used the following iterative process: First we obtained an SED fit using only the resolved photometry from \textit{HST} and VLT. Using these models, we estimated the fraction with which each of the galaxies would be contributing to the photometry in each of the unresolved bands, and used this fraction to estimate the photometry of each of the galaxies in the above-mentioned bands based on the blended photometry. Then, the SED fit was performed again using all the bands. The process was iterated until it converged. This process required only a few iterations, rapidly converging to the parameters shown in Table~\ref{Tab:hostproperties}. Figure~\ref{fig:hostsed} shows the different fits. It is worth noting that although the {\it GALEX} and {\it WISE} photometry helps to get a better constraint on the galaxy models, the results differ only slightly from those obtained just with the \textit{HST}/ACS and VLT/HAWK-I data.

\begin{table}
\caption{Properties of the host galaxy and companion derived from the SED analysis. 
}
\label{Tab:hostproperties}
\centering        
\small             
\begin{tabular}{l | c | c}    
\hline\hline                 
Property	            & Host                  & Companion	            \\  
\hline
$\chi^2$ / \# filters   & 9.1/13                & 14.6/13                  \\ \vspace{1mm}
E($B-V$) (mag)          & 0.30                  & 0.10                  \\ \vspace{1mm} 
Age (Gyr)               & $0.23_{-0.18}^{+1.30}$   & $6.12_{-3.55}^{+2.40}$   \\ \vspace{1mm}
log(Mass (M$_{\odot}$))  & $9.27_{-0.25}^{+0.28}$ & $9.95_{-0.22}^{+0.09}$  \\ \vspace{1mm}
SFR (M$_{\odot}$ yr$^{-1}$)& $9.4_{-6.4}^{+12.8}$  & $0.28_{-0.15}^{+0.51}$   \\ \vspace{1mm}
log(SSFR (yr$^{-1}$))    & $-8.3_{-0.8}^{+0.6}$ & $-10.5_{-0.4}^{+0.6}$\\ 
\hline              
\end{tabular}
\end{table}

\subsection{Molecular analysis}

\begin{table*}
\small
\renewcommand{\arraystretch}{1.2}
\begin{center}
\caption{Spectral analysis from the ALMA CO(3-2) spectroscopy. 
}
\label{table:co}
\begin{tabular}{l | c | c | c}
\hline
\hline
Property                &  Host             & Companion         &  Combined   \\
\hline
Amplitude (mJy)	        & $0.61\pm0.10$	    & $1.37\pm0.10$	    & $2.16\pm0.15$   \\
Centre (GHz)			& $242.625\pm0.010$ & $242.654\pm0.005$ & $242.646\pm0.004$   \\
FWHM (km/s)	            & $232\pm42$		& $231\pm19$		& $223\pm19$   \\
Line Flux (mJy·km/s)	& $150\pm25$		& $336\pm26$		& $512\pm41$   \\
\hline
L$_{{\rm CO}(3-2)}$ (L$_{\odot})$                   & $(2.04\pm0.34)\times10^5$ & $(4.57\pm0.35)\times10^5$ & $(6.97\pm0.56)\times10^5$   \\
L$^{\prime}_{{\rm CO}(3-2)}$ (K·km/s·pc$^2$)        & $(1.54\pm0.26)\times10^8$ & $(3.45\pm0.27)\times10^8$ & $(5.26\pm0.42)\times10^8$ \\
L$^{\prime}_{{\rm CO}(1-0)}$ (K·km/s·pc$^2$)$^{a}$    & $(2.75\pm0.46)\times10^8$ & $(6.17\pm0.48)\times10^8$ & $(9.40\pm0.75)\times10^8$ \\
\hline
SFR(CO)(M$_{\odot}$/yr)$^{b}$                         & $33\pm8$			        & $74\pm15$			        & $113\pm21$   \\
M$_{H_2}$(M$_{\odot}$)$^{c}$ [$\alpha_{CO}=5$] & $(1.38\pm0.23)\times10^9$	& $(3.08\pm0.24)\times10^9$	& $(4.70\pm0.38)\times10^9$   \\
$f_{gas}\ ^{d}$ [$\alpha_{CO}=5$] & $0.7_{-0.4}^{+0.7}$     & $0.35_{-0.12}^{+0.26}$    &  $0.44_{-0.14}^{+0.29}$ \\
M$_{H_2}$(M$_{\odot}$)$^{e}$ [$\alpha_{CO}=18.2$] & $(5.0\pm0.8)\times10^9$	& $(11.2\pm0.9)\times10^9$	& $(17.1\pm1.4)\times10^9$   \\
$f_{gas}\ ^{d}$ [$\alpha_{CO}=18.2$] & $2.7_{-1.6}^{+2.5}$     & $1.3_{-0.4}^{+1.0}$    &  $1.6_{-0.5}^{+1.0}$ \\
\hline
\hline
\end{tabular}
\tablefoot{$^{a}$Assuming L$^{\prime}_{{\rm CO}(3-2)}$/L$^{\prime}_{{\rm CO}(1-0)}$ = 0.56 \citep{Carilli2013} for star-forming galaxies; 
$^{b}$Based on the relation given by \cite{Hunt2015} for metal-poor galaxies; $^{c}$Assuming a Galactic CO-to-H$_2$ conversion factor of $\alpha_{CO}=5$ M$_{\odot}$/(K·km/s·pc$^2$); $^d$ Calculated as $f_{gas}=M_{H_2}/M_{*}$; $^{e}$Assuming a metallicity-dependent CO-to-H$_2$ conversion factor \citep{Amorin2016} of $\alpha_{CO}=18.2$ M$_{\odot}$/(K·km/s·pc$^2$).
}
\end{center}
\end{table*}

The submillimetre data at the time of our observation were dominated by the afterglow emission (for a detailed analysis of the continuum emission see Acciari et al. 2019, in press, and Misra et al. in prep.). This implies that we cannot study the continuum emission of the host galaxy.

After subtracting the afterglow continuum we are left with the CO(3-2) emission, which reveals the two galaxies identified by \textit{HST}, for which we obtained independent spectroscopy. Using the apertures shown in Fig.~\ref{Fig:alma} we extracted the spectra of the GRB host, the companion galaxy, and the combined system. The CO(3-2) lines detected in all of these spectra were used to calculate the flux density, redshift, full width half maximum (FWHM) and relative velocity of the galaxies. From the CO flux, and knowing the distance to the galaxy we can determine the luminosity of the line and, with some assumptions, determine the expected SFR for the amount of CO and the expected mass of H$_2$.
We used two assumptions for the CO-to-H$_2$ conversion factor. First we used $\alpha_{CO}=5$ M$_{\odot}$/(K·km/s·pc$^2$) (we will drop the unit in what follows) in order to enable direct comparison with previous work. Then, we used the metallicity-dependent calibration of \cite{Amorin2016}, which for the metallicity of the host results in $\alpha_{CO}=18.2$.
This spectral analysis is presented in Table~\ref{table:co}. As was already evident in Fig.~\ref{Fig:alma}, the companion galaxy is more luminous in the CO(3-2) transition, in contrast to the optical emission, where the host galaxy of GRB\,190114C dominates.

\begin{figure}[!ht]
   \centering
   \includegraphics[width=\columnwidth]{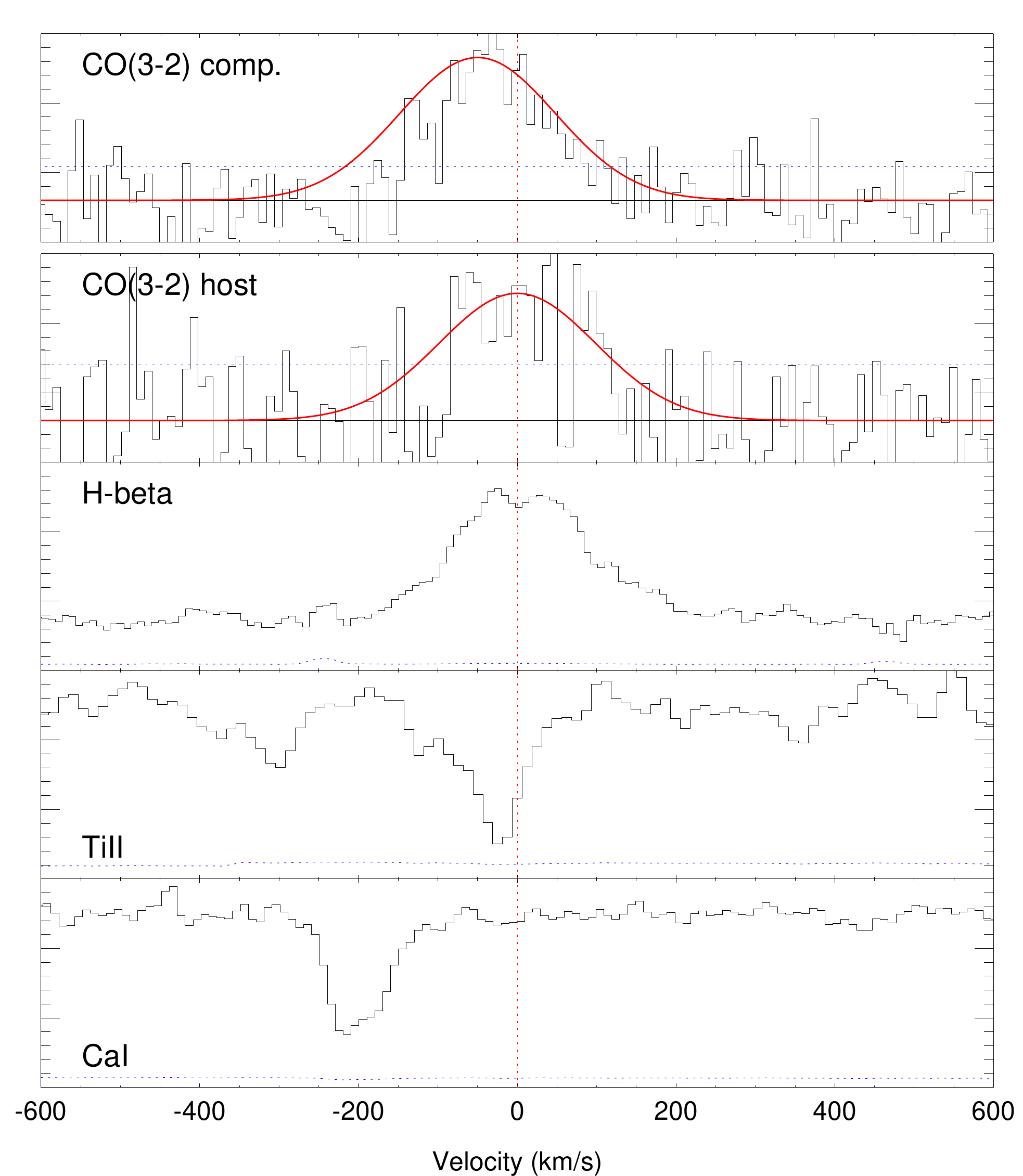}
   \caption{
   Comparison of the lines detected in the ALMA and X-shooter spectra. This plot shows how the host emission lines in both the ALMA and X-shooter spectra are coincident. However, the absorption features that are present in the spectrum are displaced and cover a much larger range in velocities than the emission, even when compared with the interacting companion galaxy, which is shown in the top panel. 
   }
         \label{Fig:lines}
   \end{figure}

The CO emission of both galaxies turns out to be almost coincident in redshift. The centres of the emission lines differ by only $50\pm20$ km/s (see Fig. \ref{Fig:lines}). This confirms that the two galaxies are, indeed, at the same distance. Together with the small projected distance between the two galaxies and the somewhat disturbed morphology shown by the GRB host, this suggests that we are seeing an interacting system.

The centre of the host galaxy and the afterglow are located close to each other, but not completely coincident. Using the ALMA data we measure a distance between them of $0\farcs094\pm0\farcs065$ equivalent to a projected distance of $530\pm370$ pc. This is larger than the $0\farcs030\pm0\farcs010$ or a projected distance of $170\pm60$ kpc derived from the HST imaging but consistent within errors, resulting from the larger uncertainty of the ALMA astrometry. In any case, a displacement of the molecular gas centroid with respect to the optical galaxy core could be a consequence of the galaxy interaction that we are observing.

\subsection{Host emission-line analysis}

We detect a number of emission lines from the host in the second epoch X-shooter spectrum (see Table~\ref{Tab:fluxes}). Among them are the usual prominent strong lines, but we also detect [S\,{\sc iii}] $\lambda$ 9071 and emission lines of [Ne\,{\sc iii}] and He\,{\sc i} $\lambda$ 5876 and 10833, the latter which are only present in young stellar populations of age $<50$ Myr \citep{GonzalezDelgado1999}. 

The absolute flux calibration of the spectrum was fine-tuned using the simultaneous GROND multiband photometric measurements (Acciari et al. 2019, in press). Several lines are affected by telluric absorption and especially the Balmer lines show signs of absorption along the line-of-sight in the host. To measure the fluxes of these lines, we mask the contaminated regions and fit the remaining part of the emission line with a Gaussian function. Fluxes of non-contaminated lines are measured by a simple integration. Errors are obtained via Monte Carlo (MC) simulation using the error spectrum and cross-checked with the r.m.s. of the spectral continuum. 
For 1000 simulated events we repeatedly added random Gaussian noise (standard deviations were taken from the error spectra) to the best-fit model and fitted the resulting spectrum by the same model. The obtained distribution of best-fit parameters was then used to compute the 1-sigma errors.
In addition, the H${\beta}$ and H${\gamma}$ lines are likely affected by underlying stellar Balmer absorption as suggested by the high stellar mass. Since even the second epoch X-shooter spectrum is dominated by afterglow emission, however, we cannot directly search for the wings of the absorption around the Balmer lines.

\begin{table}
\small
\renewcommand{\arraystretch}{1.2}
\begin{center}
\caption{Emission lines, identified in the X-shooter spectrum, and their measured fluxes corrected for Galactic extinction only. Correction of the host extinction has not been applied here due to the uncertainty in the derivation of the extinction (see text).}
\label{Tab:fluxes}
\begin{tabular}{l | c}
\hline
\hline
Line        &  $F_{\lambda}$   \\
            &   [$10^{-17}$ erg cm$^{-2}$s$^{-1}$]\\
\hline
$[$O\,{\sc ii}$]$ 3726   & $12.6 \pm 0.3$ \\
$[$O\,{\sc ii}$]$ 3729	& $16.2 \pm 0.3$ \\
$[$Ne\,{\sc iii}$]$	$^{a}$	& $2.9 \pm 0.4$ \\
H$\gamma$ $^{b}$		& $4.3 \pm 0.3$ \\
H$\beta$ $^{b}$		& $14.5 \pm 0.3$ \\
$[$O\,{\sc iii}$]$ 4959$^{a}$  & $12.8 \pm 0.3$ \\
$[$O\,{\sc iii}$]$ 5007$^{a}$  & $38.8 \pm 0.3$ \\
He\,{\sc i} 5876 		 & $2.9 \pm 0.5$ \\
$[$O\,{\sc i}$]$ 6300 	&  $2.3 \pm 0.3$ \\
H$\alpha$      & $89.7 \pm 0.8$ \\
$[$N\,{\sc ii}$]$ 6583  & $14.8 \pm 0.6$ \\
$[$S\,{\sc ii}$]$ 6717  & $8.9 \pm 0.3$ \\
$[$S\,{\sc ii}$]$ 6733  & $7.1 \pm 0.3$ \\
$[$S\,{\sc iii}$]$ 9071$^{a}$ & $10.9 \pm 0.5$ \\
He\,{\sc i} 10833       & $11.2 \pm 0.5$ \\
Pa$\gamma$ $^{c}$     & $4.9 \pm 0.5$ \\
\hline
\hline
\end{tabular}
\tablefoot{$^{a}$ Fluxes obtained by simple integration, $^{b}$ the line is affected by line-of-sight absorption in the afterglow continuum, fluxes are obtained by fitting a single Gaussian to the emission part, $^{c}$ the line suffers from residual sky emission,  the flux is obtained by using the unaffected blue wing of the line and approximating it with a Gaussian with a fixed centre.}
\end{center}
\end{table}

To determine the host-averaged attenuation we compare the Balmer decrement to the theoretical predictions for an electron temperature of $10^4\,$K and densities of  $10^2-10^4\,$cm$^{-2}$ \citep[i.e. case B recombination;][]{Osterbrock1989}. The attenuation is derived by simultaneously minimising the ratios of the H${\alpha}$, H${\beta}$ and H${\gamma}$ lines and assuming different extinction curves: the average extinction curves of the Milky Way \cite[MW,][]{Pei1992} and the Small Magellanic Cloud \cite[SMC,][]{Pei1992}, as well as the attenuation curve for starburst galaxies \citep{Calzetti2000}. Errors are obtained via MC simulation. The three different curves give slightly different results, but all point toward a relatively high value. For MW extinction we obtain E$(B-V)=0.83\pm0.03$ mag, for SMC E$(B-V)=0.76\pm0.02$ mag and a starburst attenuation curve gives E$(B-V)=0.92\pm0.03$ mag. Our analysis does not show any strong preference for a particular type of attenuation. 

The attenuation-corrected H${\alpha}$ flux is then used to estimate the star-formation rate (SFR), adopting the \cite{Kennicutt1998} relation for a \cite{Chabrier2003} IMF. The values range from $\mathrm{SFR}\sim13-25\,\mathrm{M}_{\odot}\,\mathrm{yr}^{-1}$, depending on the extinction relation used. The electron density was derived from the ratio of the [O\,{\sc ii}] lines: $n_{\rm e}=87\pm0.16$ cm$^{-3}$ \citep{Osterbrock1989}. A similar value is obtained via the ratio of the [S\,{\sc ii}] $\lambda\lambda$6727, 6732 doublet. Finally we also determine metallicities using two calibrators based on the recalibration of the O3N2 and N2 parameters with T$_e$-obtained metallicities in \citet{MarinoZ}. For O3N2 we obtain $12+\log\mathrm{(O/H)}=8.27\pm0.03$ while for N2 we get $12+\log\mathrm{(O/H)}=8.38\pm0.02$. The more reliable metallicity determination from the electron temperature cannot be applied here since the T$_e$-sensitive [O\,{\sc iii}]$ \lambda$4363 line is not detected. Note that the O3N2 parameter might suffer slightly from unaccounted stellar absorption in H$\beta$, however the values from both calibrators are consistent within errors.

\section{Discussion}

\begin{figure}[t]
\centering
\includegraphics[width=8.4cm]{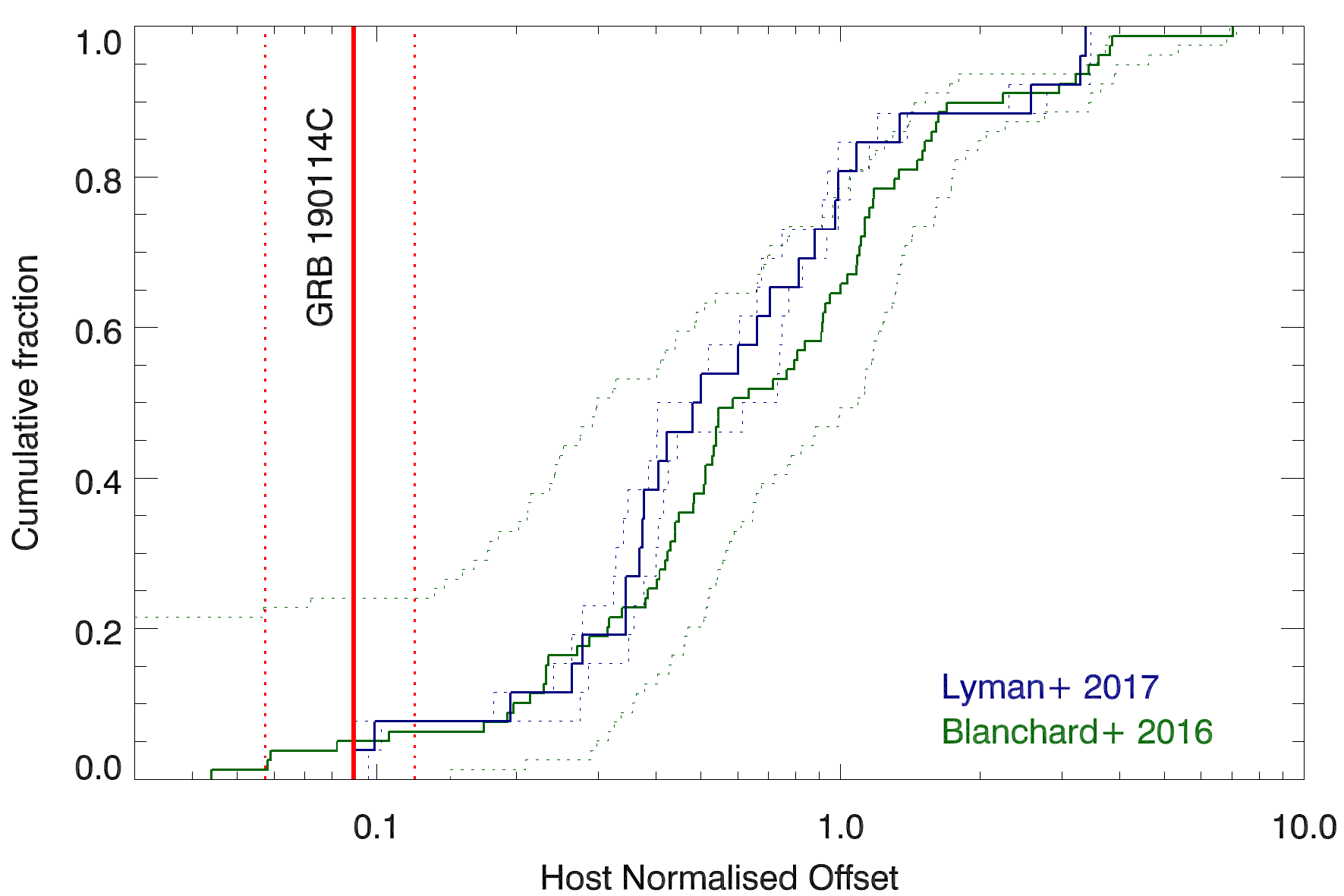}
\caption{Cumulative distribution of host normalised offsets for the samples of \citet{Blanchard2016} and \citet{Lyman2017}. The dotted lines indicate the 1-$\sigma$ uncertainties. The host normalised offset of GRB\,190114C is shown in red, and lies at the lower end of the distribution.}
\label{fig:offsets}
\end{figure}

The host of GRB\,190114C has a projected size of $1\farcs6\times0\farcs8$, and a half light radius of $r_{50}=1.90$ kpc (or an 80\% light radius of $r_{80}=3.45$ kpc). We compare this with the samples of \citet{Lyman2017} and \citet{Blanchard2016}, which used HST data of 39 and 105 GRB hosts, respectively. The \citet{Lyman2017} et al. sample comprises only observations with the F160W near infrared filter, whereas the \citet{Blanchard2016} uses whatever filter was available for each of the hosts in the HST archive, with preference towards optical data from the ACS instrument. \citet{Lyman2017} determined an average radius of $r_{50}=1.7\pm0.2$ kpc ($r_{80}=3.1\pm0.4$ kpc), which is similar to the value obtained by \citet{Blanchard2016} of $r_{50}=1.8\pm0.1$ kpc ($r_{80}\sim3$ kpc). This means that the host galaxy of our study is slightly larger than the average but well within what is typically found for GRB hosts.
Our galaxy shows a tidally disrupted morphology, with the GRB occurring within the bulge of the galaxy at $170\pm60$~pc from the galaxy centre. This offset is significantly smaller than the average one found for GRBs, which was measured to be $1.0\pm0.2$ kpc by \citet{Lyman2017} or $1.3\pm0.2$ kpc by \citet{Blanchard2016}. By examining the location of the host galaxy in a BPT diagram \citep{Baldwin1981}, we can discard the presence of a central AGN and any additional variability produced by it. This allows us to conclude that the detected emission is only related to the GRB, and the offset measured from the host galaxy core is accurate. The location of GRB\,190114C is even more outstanding when we look at the normalised offset, calculated as the ratio between the measured offset and the half light radius of the galaxy (see Fig.~\ref{fig:offsets}). The average values measured within the HST GRB host samples were of Offset/$r_{50}=0.6\pm0.1$ \citep{Lyman2017} and Offset/$r_{50}=0.7\pm0.2$. In our case we measure Offset/$r_{50}=0.09\pm0.03$.

A second galaxy is located at the same redshift, with a velocity difference of only $50\pm20$ km/s and a projected distance of $6.8\pm0.6$~kpc suggesting that the two galaxies are interacting. The mass ratio between the two galaxies is $M_{\rm comp}/M_{\rm host} = 4.8_{-3.5}^{+1.9}$ implying that the interacting system can be classified as a minor merger, which are those with a mass ratio larger than 3.0 \citep{Scudder2012}, although the uncertainty is large and could make the system also consistent with a major merger. Furthermore, the system can be classified as a close pair, which includes those galaxies with a projected separation $r_p < 30$\, kpc. Interacting systems result in an enhancement of the SFR, which is even higher in the case of major mergers and is maximised in the case of close pairs. At a distance of less than 10 kpc, \citet{Scudder2012} showed that major mergers have an average increase in SFR of a factor of two, whereas minor mergers have increases of a factor in the range 1.5--2. The interaction also results in a decrease in the metallicity of the system, as compared to non-interacting galaxies. Although this shows a weaker trend as the projected distance changes, it is also maximised for the closest interacting galaxies, where the metallicity changes by $\Delta\log(\textnormal{O/H})\sim-0.05$. 

\cite{Violino2018}, in a study of molecular gas in galaxy pairs, found that they have shorter depletion times and an enhancement of the molecular gas fraction of 0.4 dex as compared to a control sample of isolated galaxies, although being consistent with other galaxies with similarly enhanced SFR. Indeed, the host of GRB\,190114C is not molecule deficient, as has been found in other GRB hosts \citep{Michalowski2018}, and both host and companion have molecular gas fractions that are higher than those found in the control sample. We do note that \cite{Violino2018} used a CO-to-H$_2$ conversion factor that was lower than any of the ones that we used, but even considering that, both the GRB host and the companion galaxy have molecular gas fractions that are consistent with the galaxy-pair sample and larger than the control one.

The increase in SFR and the decrease in metallicity found in galaxy pairs are factors that favour the production of the stars that end up generating a long GRB on timescales of a few tens of Myr. Together with the high molecular-gas fraction found in the GRB host and companion, this leads us to believe that the influence of the interaction on the current conditions of the system have been a key factor to generate the conditions that gave rise to the progenitor of \grb.

The host is star-forming, as shown in the different methods that we applied in our analysis. The SED fit reveals a somewhat unconstrained value of $\mathrm{SFR}=9.4_{-6.4}^{+12.8}$ M$_\odot\,\mathrm{yr}^{-1}$, whereas the analysis of the strong emission lines implies a SFR ranging from 13 to 25 M$_\odot\,\mathrm{yr}^{-1}$, depending on the extinction law used. Although the values that we obtain vary, they consistently show significant star-formation activity within the host. This is not the case for the companion galaxy, for which we estimate a $\mathrm{SFR}=0.28_{-0.15}^{+0.51}$ M$_\odot\,\mathrm{yr}^{-1}$ from the SED fit.

The total SFR and molecular gas masses are known to be correlated. Directly from the CO luminosity, and assuming low metallicity ($12+\log\mathrm{(O/H)}<8.4$) we can estimate a SFR of $33\pm8$ M$_\odot\,\mathrm{yr}^{-1}$ \citep{Hunt2015} for the host galaxy, while we get a $\mathrm{SFR}=74\pm15$ M$_\odot\,\mathrm{yr}^{-1}$ for the companion. However, in the case of the companion, the low metallicity assumptions are probably not valid (although we do not have a measurement of it), leading to an overestimated star-formation rate from the CO data. In any case, this is clearly a molecule-rich galaxy.

\begin{figure}[t]
\centering
\includegraphics[width=8.35cm]{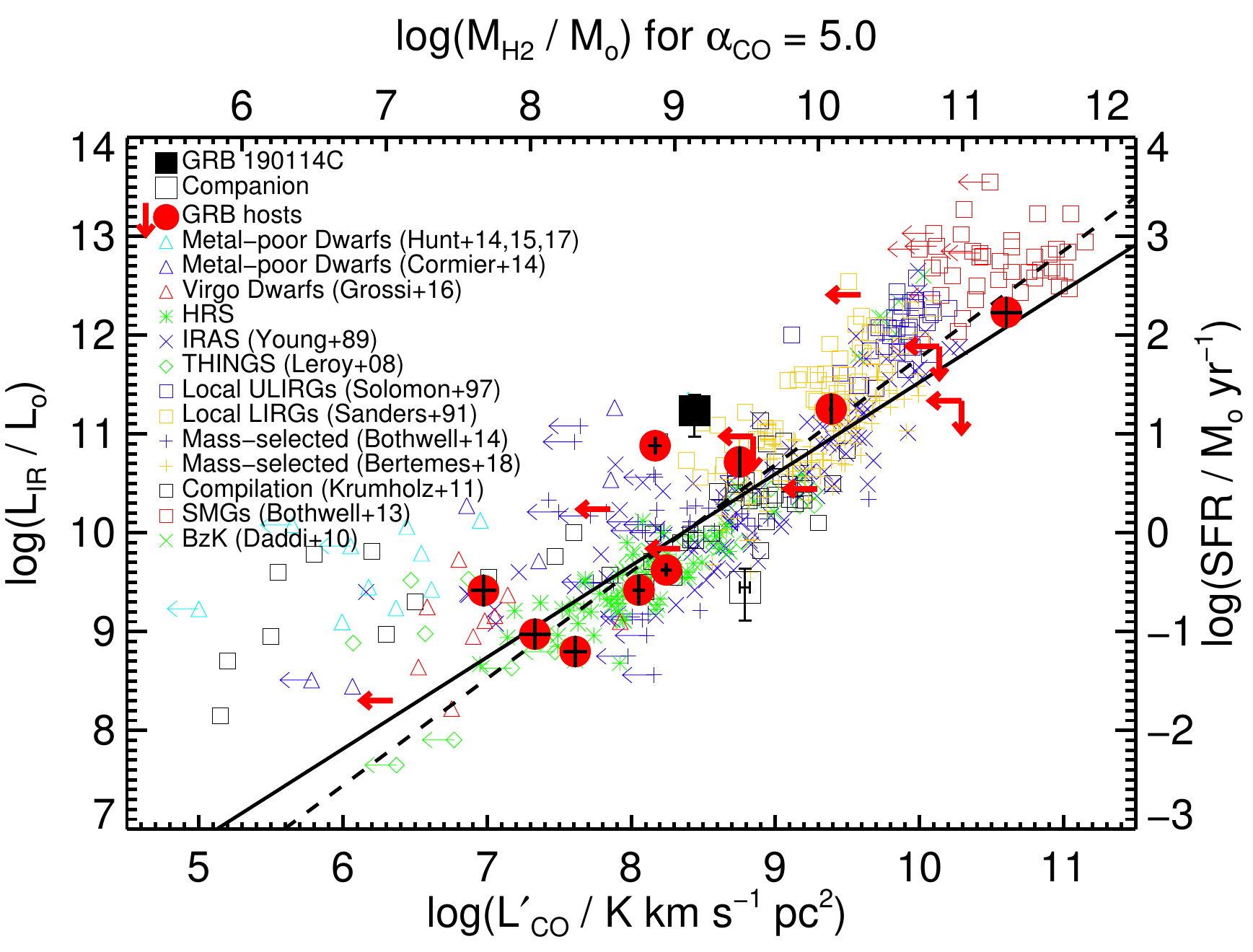}
\includegraphics[width=9cm]{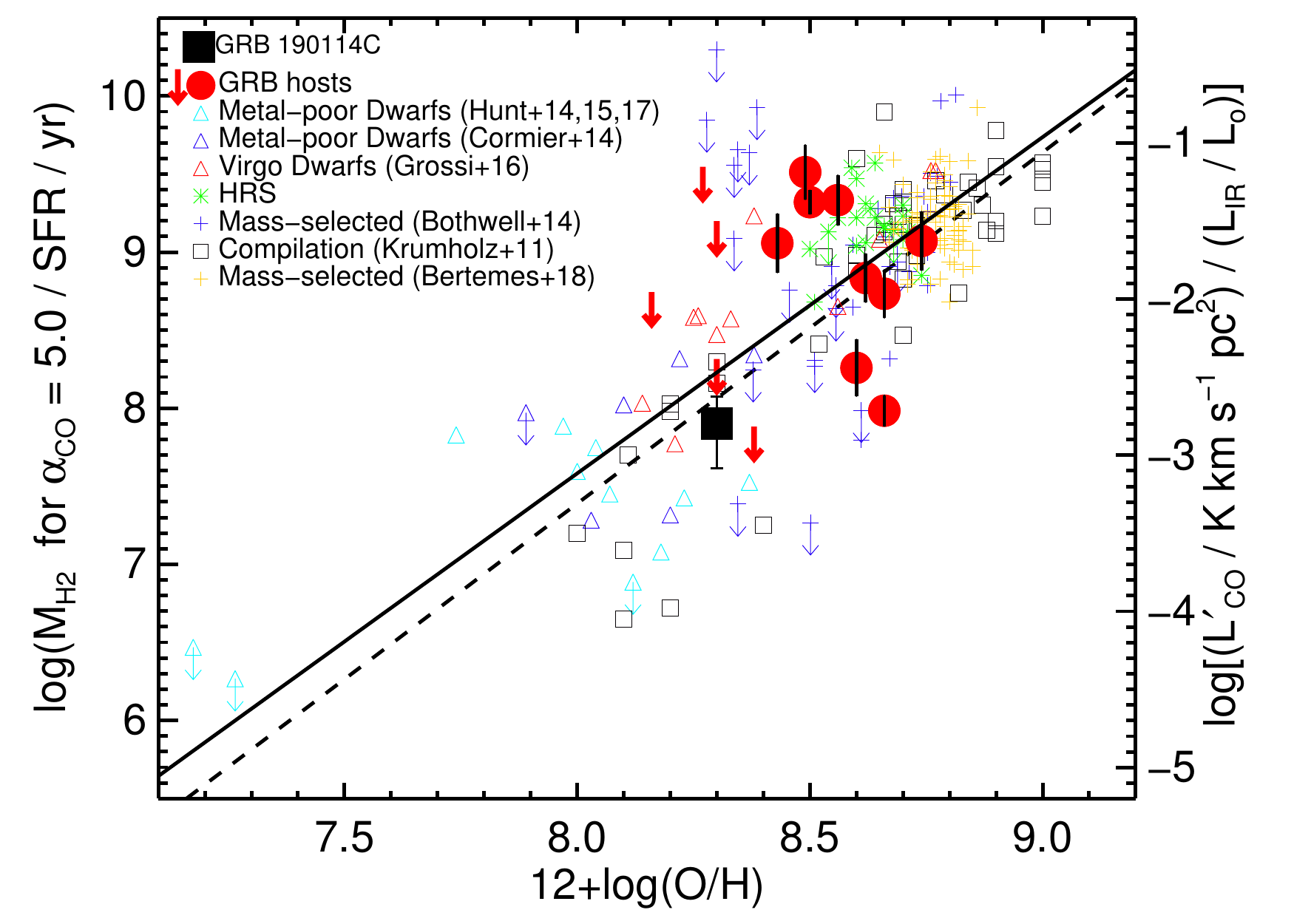}
\caption{Top: Infrared luminosity and SFR as a function of the CO luminosity, using a CO-to-H$_2$ conversion factor of $\alpha_{CO}=5$ M$_{\odot}$/(K·km/s·pc$^2$). The host of \grb~is indicated with a black, filled square, whereas the companion is an empty square. They are compared to the existing sample of GRB host galaxies and other galaxy samples. Bottom: Molecular gas depletion time (or the inverse of the star-formation efficiency), i.e. the ratio of the CO luminosity to the infrared luminosity or the corresponding molecular gas mass with the CO-to-H$_2$. In this case, only the host galaxy is shown, since the metallicity is not available for the companion, together with several galaxy samples. Adapted from \citet[][]{Michalowski2018}.}
\label{fig:co_sample}
\end{figure}

   \begin{figure}[!ht]
   \begin{minipage}{\columnwidth}
   \centering
   \includegraphics[width=\columnwidth]{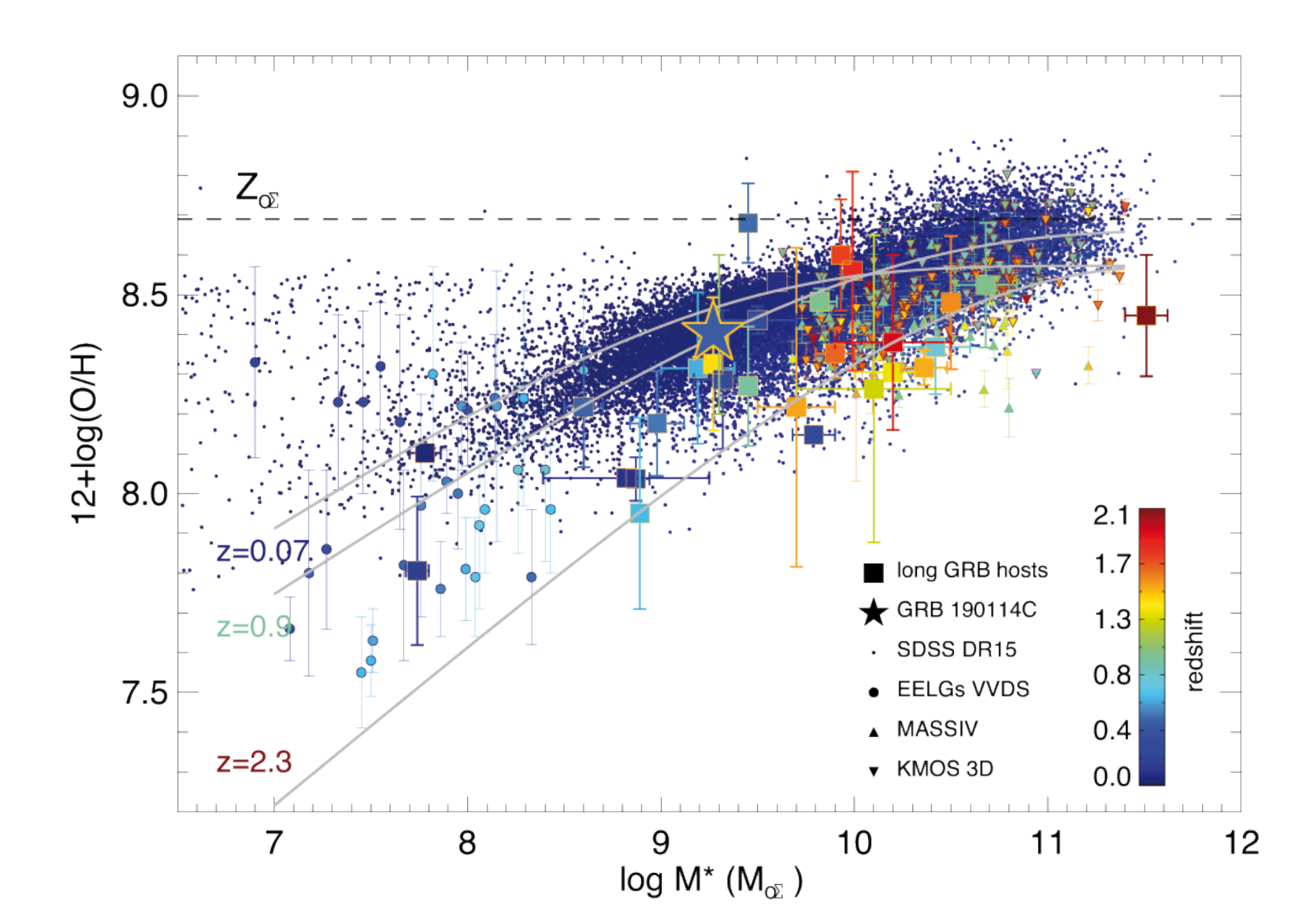}
   \caption[]{Mass-metallicity relation for the host of GRB 190114C and other long GRB hosts. As comparison samples we plot data from SDSS DR15 ($z=0-0.32$)\footnote{{\scriptsize\url{https://www.sdss.org/dr15/}}}, the KMOS3D sample ($z=0.6-2.7$, \citealt{Wuyts16}), extreme emission-line galaxies found in the VVDS ($z=0-0.9$, \citealt{Amorin2014}) as well as the MASSIVE sample ($z=0.9-2.2$, galaxies from the VVDS sample followed up by SINFONI, \citealt{Contini12})\footnote{{\scriptsize\url{ http://cosmosdb.iasf-milano.inaf.it:8080/VVDS-SINFONI/}}}. We only plot galaxies up to a redshift of 2.1 for better visualisation. All metallicities except the \cite{Amorin2014} EELG sample are derived from the N2 parameter and the \cite{MarinoZ} calibration, \cite{Amorin2014} metallicities were derived via the T$_e$ method. We also plot the MZ relations at $z=0.07$ \citep{Zahid13}, $z=0.9$ and $z=2.3$ \citep{Wuyts14}, which are also based on N2 parameter metallicities. The MZ relations were corrected for the difference in metallicity calibrators.}
         \label{Fig:MZ}
   \end{minipage}
   \end{figure}

Using a different approach based on the relation derived by \citet[][Eq.~1]{Michalowski2018}, for $\mathrm{SFR}=9.4-25$\,M$_\odot\,\mathrm{yr}^{-1}$ of the {\grb} host, we can predict $\log M_{\rm H_2}=10.05-10.49$. This is significantly larger than our estimated value $\log M_{\rm H_2}=9.14$ with $\alpha_{CO}=5$ or 9.69 with a metallicity-dependent factor \citep{Amorin2016} of $\alpha_{CO}=18.2$, so the host would be molecule-deficient, as claimed for a few other GRB hosts \citep{Hatsukade2014,Stanway2015,Michalowski2016,Michalowski2018}. On the other hand, the companion galaxy with $\mbox{SFR}=0.28$\,M$_\odot\,\mathrm{yr}^{-1}$ is expected to have $\log M_{\rm H_2}=8.44$, an order of magnitude less than the measured value. Hence, the companion is very molecule-rich compared to other galaxies with similar SFR. In the top panel of Fig.~\ref{fig:co_sample}, adapted from \citet{Michalowski2018}, we plot the two galaxies as compared to other GRB hosts and several galaxy samples \citep{Hunt2014,Hunt2015,Hunt2017,Cormier2014,Grossi2016,Young1989,Leroy2008,Solomon1997,Sanders1991,Bothwell2013,Bothwell2014,Bertemes2018,Krumholz2011,Daddi2010}. However, the  CO-to-H$_2$ conversion factor depends on metallicity \citep{Bolatto2013}. This was parametrised by \citet[][Eq.~3]{Michalowski2018} as a relation between molecular-gas depletion time $M_{\rm H_2}/\mbox{SFR}$ and metallicity. For $12+\log\mathrm{(O/H)}=8.3$ for the {\grb} host, this relation predicts $\log M_{H_2}=9.21$--$9.64$,  consistent with the value that we find (see the bottom panel of Fig.~\ref{fig:co_sample}). 

The resulting gas-to-stellar ratios of the host and the companion of 0.74 and 0.44, respectively, are within the values found for other galaxies \citep{Leroy2008,Boselli2010,Boselli2014,Bothwell2014,Michalowski2015}. The value of 2.7 derived for the host using $\alpha_{CO}=18.2$ is close to the highest measurements for other galaxies. 

   \begin{figure}[!ht]
   \begin{minipage}{\columnwidth}
   \centering
   \includegraphics[width=\columnwidth]{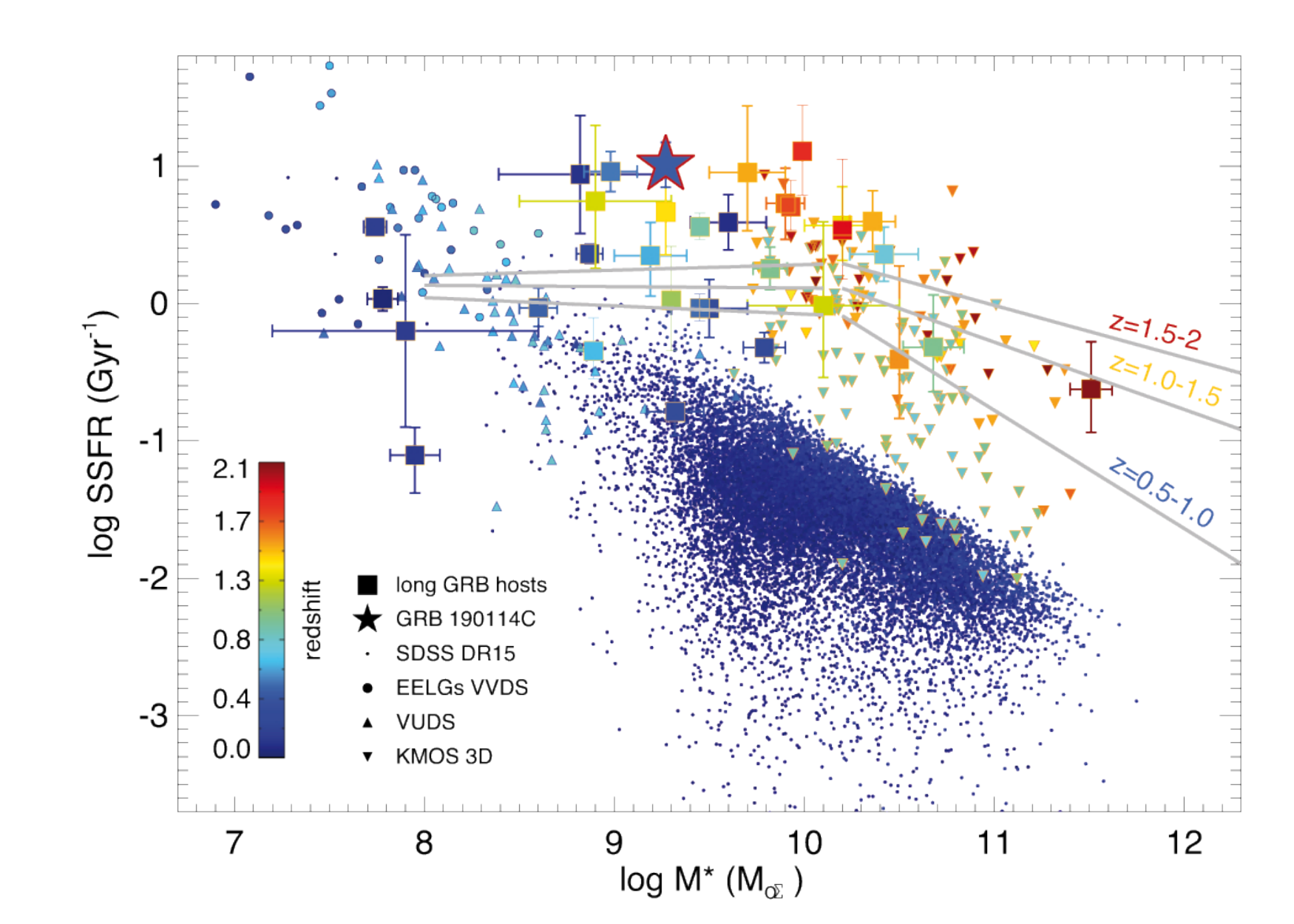}
   \caption[]{Specific SFR vs. stellar mass relation for the host of GRB 190114C and other long GRB hosts, the specific SFR was obtained by weighting it with the stellar mass. References to the comparison sampels are the same as in Fig. \ref{Fig:MZ}, we also add a sample of star-forming galaxies from the VUDS survey \citep{Calabro17} but leave out the MASSIVE sample. SFRs for the comparison samples were largely derived by SED fitting except for the KMOS3D sample. The grey lines are fits with a broken powerlaw and a cut-off mass of log $\textnormal{M}=10.2\,\textnormal{M}_\odot$ at different redshifts derived by \cite{Whitaker14}. Error bars for the comparison samples have been omitted for clarity.}
         \label{Fig:SSFR}
   \end{minipage}
   \end{figure}

To put the emission-line properties of the host into the context of other long GRB hosts we compare the mass-metallicity (MZ) relation and the SFR/SSFR-mass relation to those of other long GRBs as well as samples from the literature up to redshifts of $\sim$ 2. In Fig. \ref{Fig:MZ} we plot the MZ relation of long GRB hosts compared to galaxy samples and the fit to the MZ relation at different redshifts. In contrast to many other comparisons in the literature, we derive metallicities consistently from one single calibrator (with one exception for the VVDS EELG sample), namely the N2 parameter in the \cite{MarinoZ} calibration. The host of GRB 190114C falls slightly below the MZ relations at $z=0.07$ and even $z=0.7$, but has a relatively high mass compared to other GRB hosts at that redshift. The stellar mass is $\sim 1$ dex higher than the median value of GRB hosts at $0<z<1$ as measured for the BAT6 host sample \citep{Vergani2015} but consistent with those of $z>1$ hosts \citep{Palmerio2019}. One has to consider, however, that GRB 190114C was close to the centre of a large spiral galaxy, which usually has a negative metallicity gradient (decreasing towards the outskirts of the galaxy), while in other spiral GRB hosts the GRB occurred in the outer spiral arms \citep[see e.g.][]{Izzo19Nat}. Hence the  GRB site is likely more metal-rich than the measurements that we can make of the overall galaxy from strong-line analysis and could be close to solar, which would be exceptional for a GRB site (but see e.g. \citealt{Elliott2013,Schady2015,Michalowski18grb}). 

We also compare the SFR and SSFR (SFR weighted by the stellar mass) with other GRB hosts and field galaxies, the SSFR-mass relation is plotted in Fig \ref{Fig:SSFR}. The correlation between SFR and mass for star-forming galaxies is a broken power-law with a mass cutoff of $\log\textnormal{M}\sim10.2\,\textnormal{M}_\odot$ or possibly somewhat higher at $z>1$ \citep{Whitaker14}. Translated to the SSFR, this implies a constant SSFR  below the mass cutoff while the relation drops at higher masses. Long GRB hosts have on average higher SSFRs than e.g. field emission-line galaxies from the SDSS \citep{Japelj2016,Palmerio2019} and SSFR-M relations derived from other star-forming galaxy samples \citep[see e.g.][]{Whitaker14}. The SSFR of the host of GRB 190114C also lies above the relation for typical star-forming galaxies and has an SSFR on the upper end of the distribution of long GRBs at $z<0.5$. 

The metallicity, SFR, SSFR and stellar mass of the host galaxy of GRB\,190114C do not show any striking peculiarities with respect to other GRB hosts, except for the relatively high stellar mass for its redshift, a slightly higher SSFR as well as the atypical location of the GRB in the nucleus of the galaxy.

\section{Conclusions}

Our main conclusions are as follows:

   \begin{enumerate}
      \item High spatial-resolution imaging from HST showed that the host-galaxy system of GRB\,190114C is composed of two galaxies at a projected separation of $6.8\pm0.6$ kpc.
      \item ALMA CO(3-2) spectroscopy confirms that both galaxies are located at the same redshift, with a very low radial velocity difference, indicative that they are interacting in a plane that is almost perpendicular to the line-of-sight.
      \item These two galaxies constitute a close pair merger system. These systems are known to result in an enhancement of SFR and a decrease in metallicity, which favour the production of GRB progenitors.
      \item Both host galaxy and companion have a large molecular gas-fraction, probably induced by the interaction, which is larger than typically found in isolated galaxies.
      \item The host galaxy of GRB\,190114C has a stellar mass of log(M/M$_{\odot})=9.27_{-0.28}^{+0.25}$, larger than typical GRB hosts at this redshift and a metallicity of 12 + log(O/H) = 8.3 or 43\% solar, lower than typical field galaxies with similar mass.
      \item The host galaxy is star-forming, with a SFR that has been measured using different methods, with values in the range 3 - 25 M$_{\odot}$ yr$^{-1}$. 
      \item The presence of He\,{\sc i} emission lines in the spectrum indicates the existence of a young stellar population, in spite of the larger (although unconstrained) age derived from the SED fit.
      \item The GRB exploded within the central cluster of the host galaxy, where the density is higher, at a projected distance of $\sim170$ pc from the core. This is at the lower end of the offsets found for long GRBs, and is even more extreme when we normalise the offset with the size of the host galaxy.
      \item The fact that the GRB occurred in an extinguished line-of-sight, within the nuclear region of a galaxy, which is more massive than the average GRB host at that redshift, and has a significant molecular gas fraction is indicative of a particularly dense environment, which could have been crucial for generating the VHE photons observed by MAGIC. 
    \end{enumerate}

More detailed studies of the close local environment of the GRB will be required to determine how peculiar this event was and whether the potentially dense environment could have produced the observed VHE photons.

\begin{acknowledgements}
This paper makes use of the following ALMA data: ADS/JAO.ALMA\#2018.A.00020.T. ALMA is a partnership of ESO (representing its member states), NSF (USA) and NINS (Japan), together with NRC (Canada), MOST and ASIAA (Taiwan), and KASI (Republic of Korea), in cooperation with the Republic of Chile. The Joint ALMA Observatory is operated by ESO, AUI/NRAO and NAOJ.
CT, AdUP, DAK, MB, and LI acknowledge support from the Spanish research project AYA2017-89384-P. CT and AdUP acknowledge support from funding associated to Ram\'on y Cajal fellowships (RyC-2012-09984 and RyC-2012-09975). M.J.M.~acknowledges the support of the National Science Centre, Poland through the SONATA BIS grant 2018/30/E/ST9/00208. JPUF thanks the Carlsberg Foundation for support. KEH and PJ acknowledge support by a Project Grant (162948--051) from The Icelandic Research Fund. JS and DW are supported in part by Independent Research Fund Denmark grant DFF-7014-00017. The Cosmic Dawn Center is supported by the Danish National Research Foundation under grant No. 140. 
\end{acknowledgements}


\bibliographystyle{aa}

\begin{thebibliography}{109}
\expandafter\ifx\csname natexlab\endcsname\relax\def\natexlab#1{#1}\fi

\bibitem[{{Ajello} {et~al.}(2019){Ajello}, {Arimoto}, {Axelsson}, {Baldini},
  {Barbiellini}, {Bastieri}, {Bellazzini}, {Bhat}, {Bissaldi}, {Blandford},
  {Bonino}, {Bonnell}, {Bottacini}, {Bregeon}, {Bruel}, {Buehler}, {Cameron},
  {Caputo}, {Caraveo}, {Cavazzuti}, {Chen}, {Cheung}, {Chiaro}, {Ciprini},
  {Costantin}, {Crnogorcevic}, {Cutini}, {Dainotti}, {D'Ammand o}, {de la Torre
  Luque}, {de Palma}, {Desai}, {Desiante}, {Di Lalla}, {Di Venere}, {Fana
  Dirirsa}, {Fegan}, {Franckowiak}, {Fukazawa}, {Funk}, {Fusco}, {Gargano},
  {Gasparrini}, {Giglietto}, {Giordano}, {Giroletti}, {Green}, {Grenier},
  {Grove}, {Guiriec}, {Hays}, {Hewitt}, {Horan}, {J{\'o}hannesson}, {Kocevski},
  {Kuss}, {Latronico}, {Li}, {Longo}, {Loparco}, {Lovellette}, {Lubrano},
  {Maldera}, {Manfreda}, {Mart{\'\i}-Devesa}, {Mazziotta}, {Mereu}, {Meyer},
  {Michelson}, {Mirabal}, {Mitthumsiri}, {Mizuno}, {Monzani}, {Moretti},
  {Morselli}, {Moskalenko}, {Negro}, {Nuss}, {Ohno}, {Omodei}, {Orienti},
  {Orlando}, {Palatiello}, {Paliya}, {Paneque}, {Persic}, {Pesce-Rollins},
  {Petrosian}, {Piron}, {Poolakkil}, {Poon}, {Porter}, {Principe}, {Racusin},
  {Rain{\`o}}, {Rando}, {Razzano}, {Razzaque}, {Reimer}, {Reimer}, {Reposeur},
  {Ryde}, {Serini}, {Sgr{\`o}}, {Siskind}, {Sonbas}, {Spandre}, {Spinelli},
  {Suson}, {Tajima}, {Takahashi}, {Tak}, {Thayer}, {Torres}, {Troja},
  {Valverde}, {Veres}, {Vianello}, {von Kienlin}, {Wood}, {Yassine}, {Zhu}, \&
  {Zimmer}}]{Ajello2019}
{Ajello}, M., {Arimoto}, M., {Axelsson}, M., {et~al.} 2019, \apj, 878, 52

\bibitem[{{Amor{\'\i}n} {et~al.}(2016){Amor{\'\i}n}, {Mu{\~n}oz-Tu{\~n}{\'o}n},
  {Aguerri}, \& {Planesas}}]{Amorin2016}
{Amor{\'\i}n}, R., {Mu{\~n}oz-Tu{\~n}{\'o}n}, C., {Aguerri}, J.~A.~L., \&
  {Planesas}, P. 2016, \aap, 588, A23

\bibitem[{{Amor{\'\i}n} {et~al.}(2014){Amor{\'\i}n}, {Sommariva}, {Castellano},
  {Grazian}, {Tasca}, {Fontana}, {Pentericci}, {Cassata}, {Garilli}, {Le Brun},
  {Le F{\`e}vre}, {Maccagni}, {Thomas}, {Vanzella}, {Zamorani}, {Zucca},
  {Bardelli}, {Capak}, {Cassar{\'a}}, {Cimatti}, {Cuby}, {Cucciati}, {de la
  Torre}, {Durkalec}, {Giavalisco}, {Hathi}, {Ilbert}, {Lemaux}, {Moreau},
  {Paltani}, {Ribeiro}, {Salvato}, {Schaerer}, {Scodeggio}, {Talia},
  {Taniguchi}, {Tresse}, {Vergani}, {Wang}, {Charlot}, {Contini}, {Fotopoulou},
  {L{\'o}pez-Sanjuan}, {Mellier}, \& {Scoville}}]{Amorin2014}
{Amor{\'\i}n}, R., {Sommariva}, V., {Castellano}, M., {et~al.} 2014, \aap, 568,
  L8

\bibitem[{{Arnouts} {et~al.}(1999){Arnouts}, {Cristiani}, {Moscardini},
  {Matarrese}, {Lucchin}, {Fontana}, \& {Giallongo}}]{Arnouts1999}
{Arnouts}, S., {Cristiani}, S., {Moscardini}, L., {et~al.} 1999, \mnras, 310,
  540

\bibitem[{{Baldwin} {et~al.}(1981){Baldwin}, {Phillips}, \&
  {Terlevich}}]{Baldwin1981}
{Baldwin}, J.~A., {Phillips}, M.~M., \& {Terlevich}, R. 1981, \pasp, 93, 5

\bibitem[{{Bertemes} {et~al.}(2018){Bertemes}, {Wuyts}, {Lutz}, {F{\"o}rster
  Schreiber}, {Genzel}, {Minchin}, {Mundell}, {Rosario}, {Saintonge}, \&
  {Tacconi}}]{Bertemes2018}
{Bertemes}, C., {Wuyts}, S., {Lutz}, D., {et~al.} 2018, \mnras, 478, 1442

\bibitem[{{Bertin} \& {Arnouts}(1996)}]{Bertin1996}
{Bertin}, E. \& {Arnouts}, S. 1996, \aaps, 117, 393

\bibitem[{{Bianchi} {et~al.}(2011){Bianchi}, {Herald}, {Efremova}, {Girardi},
  {Zabot}, {Marigo}, {Conti}, \& {Shiao}}]{Bianchi2011}
{Bianchi}, L., {Herald}, J., {Efremova}, B., {et~al.} 2011, \apss, 335, 161

\bibitem[{{Blanchard} {et~al.}(2016){Blanchard}, {Berger}, \&
  {Fong}}]{Blanchard2016}
{Blanchard}, P.~K., {Berger}, E., \& {Fong}, W.-f. 2016, \apj, 817, 144

\bibitem[{{Bolatto} {et~al.}(2013){Bolatto}, {Wolfire}, \&
  {Leroy}}]{Bolatto2013}
{Bolatto}, A.~D., {Wolfire}, M., \& {Leroy}, A.~K. 2013, \araa, 51, 207

\bibitem[{{Boselli} {et~al.}(2014){Boselli}, {Cortese}, \&
  {Boquien}}]{Boselli2014}
{Boselli}, A., {Cortese}, L., \& {Boquien}, M. 2014, \aap, 564, A65

\bibitem[{{Boselli} {et~al.}(2010){Boselli}, {Eales}, {Cortese}, {Bendo},
  {Chanial}, {Buat}, {Davies}, {Auld}, {Rigby}, {Baes}, {Barlow}, {Bock},
  {Bradford}, {Castro-Rodriguez}, {Charlot}, {Clements}, {Cormier}, {Dwek},
  {Elbaz}, {Galametz}, {Galliano}, {Gear}, {Glenn}, {Gomez}, {Griffin}, {Hony},
  {Isaak}, {Levenson}, {Lu}, {Madden}, {O'Halloran}, {Okamura}, {Oliver},
  {Page}, {Panuzzo}, {Papageorgiou}, {Parkin}, {Perez-Fournon}, {Pohlen},
  {Rangwala}, {Roussel}, {Rykala}, {Sacchi}, {Sauvage}, {Schulz}, {Schirm},
  {Smith}, {Spinoglio}, {Stevens}, {Symeonidis}, {Vaccari}, {Vigroux},
  {Wilson}, {Wozniak}, {Wright}, \& {Zeilinger}}]{Boselli2010}
{Boselli}, A., {Eales}, S., {Cortese}, L., {et~al.} 2010, \pasp, 122, 261

\bibitem[{{Bothwell} {et~al.}(2013){Bothwell}, {Smail}, {Chapman}, {Genzel},
  {Ivison}, {Tacconi}, {Alaghband -Zadeh}, {Bertoldi}, {Blain}, {Casey}, {Cox},
  {Greve}, {Lutz}, {Neri}, {Omont}, \& {Swinbank}}]{Bothwell2013}
{Bothwell}, M.~S., {Smail}, I., {Chapman}, S.~C., {et~al.} 2013, \mnras, 429,
  3047

\bibitem[{{Bothwell} {et~al.}(2014){Bothwell}, {Wagg}, {Cicone}, {Maiolino},
  {M{\o}ller}, {Aravena}, {De Breuck}, {Peng}, {Espada}, {Hodge},
  {Impellizzeri}, {Mart{\'\i}n}, {Riechers}, \& {Walter}}]{Bothwell2014}
{Bothwell}, M.~S., {Wagg}, J., {Cicone}, C., {et~al.} 2014, \mnras, 445, 2599

\bibitem[{{B{\"o}ttcher} \& {Dermer}(1998)}]{Bottcher1998}
{B{\"o}ttcher}, M. \& {Dermer}, C.~D. 1998, \apjl, 499, L131

\bibitem[{{Bruzual} \& {Charlot}(2003)}]{Bruzual2003}
{Bruzual}, G. \& {Charlot}, S. 2003, \mnras, 344, 1000

\bibitem[{{Calabr{\`o}} {et~al.}(2017){Calabr{\`o}}, {Amor{\'\i}n}, {Fontana},
  {P{\'e}rez-Montero}, {Lemaux}, {Ribeiro}, {Bardelli}, {Castellano},
  {Contini}, {De Barros}, {Garilli}, {Grazian}, {Guaita}, {Hathi}, {Koekemoer},
  {Le F{\`e}vre}, {Maccagni}, {Pentericci}, {Schaerer}, {Talia}, {Tasca}, \&
  {Zucca}}]{Calabro17}
{Calabr{\`o}}, A., {Amor{\'\i}n}, R., {Fontana}, A., {et~al.} 2017, \aap, 601,
  A95

\bibitem[{{Calzetti} {et~al.}(2000){Calzetti}, {Armus}, {Bohlin}, {Kinney},
  {Koornneef}, \& {Storchi-Bergmann}}]{Calzetti2000}
{Calzetti}, D., {Armus}, L., {Bohlin}, R.~C., {et~al.} 2000, \apj, 533, 682

\bibitem[{{Carilli} \& {Walter}(2013)}]{Carilli2013}
{Carilli}, C.~L. \& {Walter}, F. 2013, \araa, 51, 105

\bibitem[{{Castro-Tirado} {et~al.}(2019){Castro-Tirado}, {Hu},
  {Fernandez-Garcia}, {Valeev}, {Sokolov}, {Guziy}, {Oates}, {Jeong}, {Pandey},
  {Carrasco}, \& {Reverte-Paya}}]{Castro-Tirado2019}
{Castro-Tirado}, A.~J., {Hu}, Y., {Fernandez-Garcia}, E., {et~al.} 2019, GRB
  Coordinates Network, 23708

\bibitem[{{Chabrier}(2003)}]{Chabrier2003}
{Chabrier}, G. 2003, \pasp, 115, 763

\bibitem[{{Contini} {et~al.}(2012){Contini}, {Garilli}, {Le F{\`e}vre},
  {Kissler-Patig}, {Amram}, {Epinat}, {Moultaka}, {Paioro}, {Queyrel}, {Tasca},
  {Tresse}, {Vergani}, {L{\'o}pez-Sanjuan}, \& {Perez-Montero}}]{Contini12}
{Contini}, T., {Garilli}, B., {Le F{\`e}vre}, O., {et~al.} 2012, \aap, 539, A91

\bibitem[{{Cormier} {et~al.}(2014){Cormier}, {Madden}, {Lebouteiller}, {Hony},
  {Aalto}, {Costagliola}, {Hughes}, {R{\'e}my-Ruyer}, {Abel}, {Bayet},
  {Bigiel}, {Cannon}, {Cumming}, {Galametz}, {Galliano}, {Viti}, \&
  {Wu}}]{Cormier2014}
{Cormier}, D., {Madden}, S.~C., {Lebouteiller}, V., {et~al.} 2014, \aap, 564,
  A121

\bibitem[{{Daddi} {et~al.}(2010){Daddi}, {Bournaud}, {Walter}, {Dannerbauer},
  {Carilli}, {Dickinson}, {Elbaz}, {Morrison}, {Riechers}, {Onodera}, {Salmi},
  {Krips}, \& {Stern}}]{Daddi2010}
{Daddi}, E., {Bournaud}, F., {Walter}, F., {et~al.} 2010, \apj, 713, 686

\bibitem[{{de Naurois}(2019)}]{deNaurois2019GCN}
{de Naurois}. 2019, GRB Coordinates Network, 25566, 1

\bibitem[{{de Ugarte Postigo} {et~al.}(2019){de Ugarte Postigo}, {Kann},
  {Thoene}, \& {Izzo}}]{deUgarte2019}
{de Ugarte Postigo}, A., {Kann}, D.~A., {Thoene}, C.~C., \& {Izzo}, L. 2019,
  GRB Coordinates Network, 23692

\bibitem[{{de Ugarte Postigo} {et~al.}(2012){de Ugarte Postigo}, {Lundgren},
  {Mart{\'\i}n}, {Garcia-Appadoo}, {de Gregorio Monsalvo}, {Peck},
  {Micha{\l}owski}, {Th{\"o}ne}, {Campana}, {Gorosabel}, {Tanvir}, {Wiersema},
  {Castro-Tirado}, {Schulze}, {De Breuck}, {Petitpas}, {Hjorth}, {Jakobsson},
  {Covino}, {Fynbo}, {Winters}, {Bremer}, {Levan}, {Llorente},
  {S{\'a}nchez-Ram{\'\i}rez}, {Tello}, \& {Salvaterra}}]{deugarte2012a}
{de Ugarte Postigo}, A., {Lundgren}, A., {Mart{\'\i}n}, S., {et~al.} 2012,
  \aap, 538, A44

\bibitem[{{de Ugarte Postigo} {et~al.}(2018){de Ugarte Postigo}, {Th{\"o}ne},
  {Bolmer}, {Schulze}, {Mart{\'\i}n}, {Kann}, {D'Elia}, {Selsing},
  {Martin-Carrillo}, {Perley}, {Kim}, {Izzo}, {S{\'a}nchez-Ram{\'\i}rez},
  {Guidorzi}, {Klotz}, {Wiersema}, {Bauer}, {Bensch}, {Campana}, {Cano},
  {Covino}, {Coward}, {De Cia}, {de Gregorio-Monsalvo}, {De Pasquale}, {Fynbo},
  {Greiner}, {Gomboc}, {Hanlon}, {Hansen}, {Hartmann}, {Heintz}, {Jakobsson},
  {Kobayashi}, {Malesani}, {Martone}, {Meintjes}, {Micha{\l}owski}, {Mundell},
  {Murphy}, {Oates}, {Salmon}, {van Soelen}, {Tanvir}, {Turpin}, {Xu}, \&
  {Zafar}}]{deugarte2018}
{de Ugarte Postigo}, A., {Th{\"o}ne}, C.~C., {Bolmer}, J., {et~al.} 2018, \aap,
  620, A119

\bibitem[{{Dichiara} {et~al.}(2019){Dichiara}, {Bernardini}, {Burrows},
  {D'Avanzo}, {Gronwall}, {Gropp}, {Kennea}, {Klingler}, {Krimm}, \&
  {Kuin}}]{Dichiara2019GCN}
{Dichiara}, S., {Bernardini}, M.~G., {Burrows}, D.~N., {et~al.} 2019, GRB
  Coordinates Network, 25552, 1

\bibitem[{{Elliott} {et~al.}(2013){Elliott}, {Kr{\"u}hler}, {Greiner},
  {Savaglio}, {Olivares}, {Rau}, {de Ugarte Postigo},
  {S{\'a}nchez-Ram{\'\i}rez}, {Wiersema}, \& {Schady}}]{Elliott2013}
{Elliott}, J., {Kr{\"u}hler}, T., {Greiner}, J., {et~al.} 2013, \aap, 556, A23

\bibitem[{{Fan} {et~al.}(2008){Fan}, {Piran}, {Narayan}, \& {Wei}}]{Fan2008}
{Fan}, Y.-Z., {Piran}, T., {Narayan}, R., \& {Wei}, D.-M. 2008, \mnras, 384,
  1483

\bibitem[{{Fruchter} {et~al.}(2006){Fruchter}, {Levan}, {Strolger},
  {Vreeswijk}, {Thorsett}, {Bersier}, {Burud}, {Castro Cer{\'o}n},
  {Castro-Tirado}, {Conselice}, {Dahlen}, {Ferguson}, {Fynbo}, {Garnavich},
  {Gibbons}, {Gorosabel}, {Gull}, {Hjorth}, {Holland}, {Kouveliotou}, {Levay},
  {Livio}, {Metzger}, {Nugent}, {Petro}, {Pian}, {Rhoads}, {Riess}, {Sahu},
  {Smette}, {Tanvir}, {Wijers}, \& {Woosley}}]{Fruchter06}
{Fruchter}, A.~S., {Levan}, A.~J., {Strolger}, L., {et~al.} 2006, \nat, 441,
  463

\bibitem[{{Gehrels} {et~al.}(2004){Gehrels}, {Chincarini}, {Giommi}, {Mason},
  {Nousek}, {Wells}, {White}, {Barthelmy}, {Burrows}, \&
  {Cominsky}}]{Gehrels2004}
{Gehrels}, N., {Chincarini}, G., {Giommi}, P., {et~al.} 2004, \apj, 611, 1005

\bibitem[{{Gonz{\'a}lez Delgado} {et~al.}(1999){Gonz{\'a}lez Delgado},
  {Leitherer}, \& {Heckman}}]{GonzalezDelgado1999}
{Gonz{\'a}lez Delgado}, R.~M., {Leitherer}, C., \& {Heckman}, T.~M. 1999,
  \apjs, 125, 489

\bibitem[{{Gropp} {et~al.}(2019){Gropp}, {Kennea}, {Klingler}, {Krimm},
  {LaPorte}, {Lien}, {Moss}, {Palmer}, {Sbarufatti}, \& {Siegel}}]{Gropp2019}
{Gropp}, J.~D., {Kennea}, J.~A., {Klingler}, N.~J., {et~al.} 2019, GRB
  Coordinates Network, 23688

\bibitem[{{Grossi} {et~al.}(2016){Grossi}, {Corbelli}, {Bizzocchi},
  {Giovanardi}, {Bomans}, {Coelho}, {De Looze}, {Gon{\c{c}}alves}, {Hunt},
  {Leonardo}, {Madden}, {Men{\'e}ndez-Delmestre}, {Pappalardo}, \&
  {Riguccini}}]{Grossi2016}
{Grossi}, M., {Corbelli}, E., {Bizzocchi}, L., {et~al.} 2016, \aap, 590, A27

\bibitem[{{Hatsukade} {et~al.}(2019){Hatsukade}, {Hashimoto}, {Kohno},
  {Nakanishi}, {Ohta}, {Niino}, {Tamura}, \& {T{\'o}th}}]{Hatsukade2019}
{Hatsukade}, B., {Hashimoto}, T., {Kohno}, K., {et~al.} 2019, \apj, 876, 91

\bibitem[{{Hatsukade} {et~al.}(2011){Hatsukade}, {Kohno}, {Endo}, {Nakanishi},
  \& {Ohta}}]{Hatsukade2011}
{Hatsukade}, B., {Kohno}, K., {Endo}, A., {Nakanishi}, K., \& {Ohta}, K. 2011,
  \apj, 738, 33

\bibitem[{{Hatsukade} {et~al.}(2014){Hatsukade}, {Ohta}, {Endo}, {Nakanishi},
  {Tamura}, {Hashimoto}, \& {Kohno}}]{Hatsukade2014}
{Hatsukade}, B., {Ohta}, K., {Endo}, A., {et~al.} 2014, \nat, 510, 247

\bibitem[{{Heintz} {et~al.}(2019){Heintz}, {Fynbo}, {Jakobsson}, {Xu},
  {Perley}, {Malesani}, \& {Viuho}}]{Heintz2019GCN}
{Heintz}, K.~E., {Fynbo}, J. P.~U., {Jakobsson}, P., {et~al.} 2019, GRB
  Coordinates Network, 25563, 1

\bibitem[{{Hjorth} {et~al.}(2012){Hjorth}, {Malesani}, {Jakobsson}, {Jaunsen},
  {Fynbo}, {Gorosabel}, {Kr{\"u}hler}, {Levan}, {Micha{\l}owski}, \&
  {Milvang-Jensen}}]{Hjorth2012}
{Hjorth}, J., {Malesani}, D., {Jakobsson}, P., {et~al.} 2012, \apj, 756, 187

\bibitem[{{Hunt} {et~al.}(2015){Hunt}, {Garc{\'\i}a-Burillo}, {Casasola},
  {Caselli}, {Combes}, {Henkel}, {Lundgren}, {Maiolino}, {Menten}, \&
  {Testi}}]{Hunt2015}
{Hunt}, L.~K., {Garc{\'\i}a-Burillo}, S., {Casasola}, V., {et~al.} 2015, \aap,
  583, A114

\bibitem[{{Hunt} {et~al.}(2014){Hunt}, {Palazzi}, {Micha{\l}owski}, {Rossi},
  {Savaglio}, {Basa}, {Berta}, {Bianchi}, {Covino}, {D'Elia}, {Ferrero},
  {G{\"o}tz}, {Greiner}, {Klose}, {Le Borgne}, {Le Floc'h}, {Pian},
  {Piranomonte}, {Schady}, \& {Vergani}}]{Hunt2014}
{Hunt}, L.~K., {Palazzi}, E., {Micha{\l}owski}, M.~J., {et~al.} 2014, \aap,
  565, A112

\bibitem[{{Hunt} {et~al.}(2017){Hunt}, {Wei{\ss}}, {Henkel}, {Combes},
  {Garc{\'\i}a-Burillo}, {Casasola}, {Caselli}, {Lundgren}, {Maiolino},
  {Menten}, \& {Testi}}]{Hunt2017}
{Hunt}, L.~K., {Wei{\ss}}, A., {Henkel}, C., {et~al.} 2017, \aap, 606, A99

\bibitem[{{Ilbert} {et~al.}(2006){Ilbert}, {Arnouts}, {McCracken},
  {Bolzonella}, {Bertin}, {Le F{\`e}vre}, {Mellier}, {Zamorani}, {Pell{\`o}},
  {Iovino}, {Tresse}, {Le Brun}, {Bottini}, {Garilli}, {Maccagni}, {Picat},
  {Scaramella}, {Scodeggio}, {Vettolani}, {Zanichelli}, {Adami}, {Bardelli},
  {Cappi}, {Charlot}, {Ciliegi}, {Contini}, {Cucciati}, {Foucaud}, {Franzetti},
  {Gavignaud}, {Guzzo}, {Marano}, {Marinoni}, {Mazure}, {Meneux}, {Merighi},
  {Paltani}, {Pollo}, {Pozzetti}, {Radovich}, {Zucca}, {Bondi}, {Bongiorno},
  {Busarello}, {de La Torre}, {Gregorini}, {Lamareille}, {Mathez}, {Merluzzi},
  {Ripepi}, {Rizzo}, \& {Vergani}}]{Ilbert2006}
{Ilbert}, O., {Arnouts}, S., {McCracken}, H.~J., {et~al.} 2006, \aap, 457, 841

\bibitem[{{Inoue} {et~al.}(2013){Inoue}, {Granot}, {O'Brien}, {Asano},
  {Bouvier}, {Carosi}, {Connaughton}, {Garczarczyk}, {Gilmore}, {Hinton},
  {Inoue}, {Ioka}, {Kakuwa}, {Markoff}, {Murase}, {Osborne}, {Otte},
  {Starling}, {Tajima}, {Teshima}, {Toma}, {Wagner}, {Wijers}, {Williams},
  {Yamamoto}, {Yamazaki}, \& {CTA Consortium}}]{Inoue2013}
{Inoue}, S., {Granot}, J., {O'Brien}, P.~T., {et~al.} 2013, Astroparticle
  Physics, 43, 252

\bibitem[{{Izzo} {et~al.}(2019){Izzo}, {de Ugarte Postigo}, {Maeda},
  {Th{\"o}ne}, {Kann}, {Della Valle}, {Sagues Carracedo}, {Micha{\l}owski},
  {Schady}, {Schmidl}, {Selsing}, {Starling}, {Suzuki}, {Bensch}, {Bolmer},
  {Campana}, {Cano}, {Covino}, {Fynbo}, {Hartmann}, {Heintz}, {Hjorth},
  {Japelj}, {Kami{\'n}ski}, {Kaper}, {Kouveliotou}, {Kru{\.Z}y{\'n}ski},
  {Kwiatkowski}, {Leloudas}, {Levan}, {Malesani}, {Micha{\l}owski},
  {Piranomonte}, {Pugliese}, {Rossi}, {S{\'a}nchez-Ram{\'\i}rez}, {Schulze},
  {Steeghs}, {Tanvir}, {Ulaczyk}, {Vergani}, \& {Wiersema}}]{Izzo19Nat}
{Izzo}, L., {de Ugarte Postigo}, A., {Maeda}, K., {et~al.} 2019, \nat, 565, 324

\bibitem[{{Japelj} {et~al.}(2016){Japelj}, {Vergani}, {Salvaterra}, {D'Avanzo},
  {Mannucci}, {Fernandez-Soto}, {Boissier}, {Hunt}, {Atek},
  {Rodr{\'{\i}}guez-Mu{\~n}oz}, {Scodeggio}, {Cristiani}, {Le Floc'h},
  {Flores}, {Gallego}, {Ghirlanda}, {Gomboc}, {Hammer}, {Perley}, {Pescalli},
  {Petitjean}, {Puech}, {Rafelski}, \& {Tagliaferri}}]{Japelj2016}
{Japelj}, J., {Vergani}, S.~D., {Salvaterra}, R., {et~al.} 2016, \aap, 590,
  A129

\bibitem[{{Japelj} {et~al.}(2018){Japelj}, {Vergani}, {Salvaterra}, {Renzo},
  {Zapartas}, {de Mink}, {Kaper}, \& {Zibetti}}]{Japelj2018}
{Japelj}, J., {Vergani}, S.~D., {Salvaterra}, R., {et~al.} 2018, \aap, 617,
  A105

\bibitem[{{Kann} {et~al.}(2019){Kann}, {Thoene}, {Selsing}, {Izzo}, {de Ugarte
  Postigo}, {Pugliese}, {Sbarufatti}, {Heintz}, {D'Elia}, {Covino}, {Wiersema},
  {Perley}, {Vergani}, {Fynbo}, {Watson}, {Tanvir}, {Hartmann}, {Xu},
  {Schulze}, \& {Bolmer}}]{Kann2019}
{Kann}, D.~A., {Thoene}, C.~C., {Selsing}, J., {et~al.} 2019, GRB Coordinates
  Network, 23710

\bibitem[{{Kelly} {et~al.}(2014){Kelly}, {Filippenko}, {Modjaz}, \&
  {Kocevski}}]{Kelly2014}
{Kelly}, P.~L., {Filippenko}, A.~V., {Modjaz}, M., \& {Kocevski}, D. 2014,
  \apj, 789, 23

\bibitem[{{Kelly} {et~al.}(2008){Kelly}, {Kirshner}, \& {Pahre}}]{Kelly2008}
{Kelly}, P.~L., {Kirshner}, R.~P., \& {Pahre}, M. 2008, \apj, 687, 1201

\bibitem[{{Kennicutt}(1998)}]{Kennicutt1998}
{Kennicutt}, Jr., R.~C. 1998, \araa, 36, 189

\bibitem[{{Krimm} {et~al.}(2019){Krimm}, {Barthelmy}, {Cummings}, {Gropp},
  {Lien}, {Markwardt}, {Palmer}, {Sakamoto}, {Stamatikos}, \&
  {Ukwatta}}]{Krimm2019}
{Krimm}, H.~A., {Barthelmy}, S.~D., {Cummings}, J.~R., {et~al.} 2019, GRB
  Coordinates Network, 23724

\bibitem[{{Kron}(1980)}]{Kron1980}
{Kron}, R.~G. 1980, \apjs, 43, 305

\bibitem[{{Kr{\"u}hler} {et~al.}(2015){Kr{\"u}hler}, {Malesani}, {Fynbo},
  {Hartoog}, {Hjorth}, {Jakobsson}, {Perley}, {Rossi}, {Schady}, {Schulze},
  {Tanvir}, {Vergani}, {Wiersema}, {Afonso}, {Bolmer}, {Cano}, {Covino},
  {D'Elia}, {de Ugarte Postigo}, {Filgas}, {Friis}, {Graham}, {Greiner},
  {Goldoni}, {Gomboc}, {Hammer}, {Japelj}, {Kann}, {Kaper}, {Klose}, {Levan},
  {Leloudas}, {Milvang-Jensen}, {Nicuesa Guelbenzu}, {Palazzi}, {Pian},
  {Piranomonte}, {S{\'a}nchez-Ram{\'\i}rez}, {Savaglio}, {Selsing},
  {Tagliaferri}, {Vreeswijk}, {Watson}, \& {Xu}}]{Kruehler2015}
{Kr{\"u}hler}, T., {Malesani}, D., {Fynbo}, J.~P.~U., {et~al.} 2015, \aap, 581,
  A125

\bibitem[{{Krumholz} {et~al.}(2011){Krumholz}, {Leroy}, \&
  {McKee}}]{Krumholz2011}
{Krumholz}, M.~R., {Leroy}, A.~K., \& {McKee}, C.~F. 2011, \apj, 731, 25

\bibitem[{{Laskar} {et~al.}(2018){Laskar}, {Alexander}, {Berger}, {Guidorzi},
  {Margutti}, {Fong}, {Kilpatrick}, {Milne}, {Drout}, {Mundell}, {Kobayashi},
  {Lunnan}, {Barniol Duran}, {Menten}, {Ioka}, \& {Williams}}]{Laskar2018}
{Laskar}, T., {Alexander}, K.~D., {Berger}, E., {et~al.} 2018, \apj, 862, 94

\bibitem[{{Laskar} {et~al.}(2019){Laskar}, {Alexander}, {Gill}, {Granot},
  {Berger}, {Mundell}, {Barniol Duran}, {Bolmer}, {Duffell}, {van Eerten},
  {Fong}, {Kobayashi}, {Margutti}, \& {Schady}}]{Laskar2019}
{Laskar}, T., {Alexander}, K.~D., {Gill}, R., {et~al.} 2019, \apjl, 878, L26

\bibitem[{{Leroy} {et~al.}(2008){Leroy}, {Walter}, {Brinks}, {Bigiel}, {de
  Blok}, {Madore}, \& {Thornley}}]{Leroy2008}
{Leroy}, A.~K., {Walter}, F., {Brinks}, E., {et~al.} 2008, \aj, 136, 2782

\bibitem[{{Lipunov} {et~al.}(2019){Lipunov}, {Tyurina}, {Kuznetsov},
  {Gorbovskoy}, {Kornilov}, {Kushinov}, {Balanutsa}, {Vlasenko}, {Vladimirov},
  {Kuznetsov}, {Chasovnikov}, {Rebolo}, {Serra}, {Lodieu}, {Israelian},
  {Suarez-Andres}, {Buckley}, {Potter}, {Yurkov}, {Gabovich}, {Sergienko},
  {Kobcev}, {Tlatov}, {Senik}, {Parhomenko}, {Dormidontov}, {Gres}, {Budnev},
  {Yazev}, {Chuvalaev}, {Poleshchuk}, {Podesta}, {Lopez}, {Francile},
  {Podesta}, \& {Levato}}]{Lipunov2019}
{Lipunov}, V., {Tyurina}, N., {Kuznetsov}, A., {et~al.} 2019, GRB Coordinates
  Network, 23693

\bibitem[{{Lyman} {et~al.}(2017){Lyman}, {Levan}, {Tanvir}, {Fynbo}, {McGuire},
  {Perley}, {Angus}, {Bloom}, {Conselice}, {Fruchter}, {Hjorth}, {Jakobsson},
  \& {Starling}}]{Lyman2017}
{Lyman}, J.~D., {Levan}, A.~J., {Tanvir}, N.~R., {et~al.} 2017, \mnras, 467,
  1795

\bibitem[{{MacKenty} {et~al.}(2010){MacKenty}, {Kimble}, {O'Connell}, \&
  {Townsend}}]{MacKenty2010}
{MacKenty}, J.~W., {Kimble}, R.~A., {O'Connell}, R.~W., \& {Townsend}, J.~A.
  2010, in Society of Photo-Optical Instrumentation Engineers (SPIE) Conference
  Series, Vol. 7731, \procspie, 77310Z

\bibitem[{{Marino} {et~al.}(2013){Marino}, {Rosales-Ortega}, {S{\'a}nchez},
  {Gil de Paz}, {V{\'{\i}}lchez}, {Miralles-Caballero}, {Kehrig},
  {P{\'e}rez-Montero}, {Stanishev}, {Iglesias-P{\'a}ramo}, {D{\'{\i}}az},
  {Castillo-Morales}, {Kennicutt}, {L{\'o}pez-S{\'a}nchez}, {Galbany},
  {Garc{\'{\i}}a-Benito}, {Mast}, {Mendez-Abreu}, {Monreal-Ibero}, {Husemann},
  {Walcher}, {Garc{\'{\i}}a-Lorenzo}, {Masegosa}, {Del Olmo Orozco},
  {Mour{\~a}o}, {Ziegler}, {Moll{\'a}}, {Papaderos},
  {S{\'a}nchez-Bl{\'a}zquez}, {Gonz{\'a}lez Delgado}, {Falc{\'o}n-Barroso},
  {Roth}, {van de Ven}, \& {Califa Team}}]{MarinoZ}
{Marino}, R.~A., {Rosales-Ortega}, F.~F., {S{\'a}nchez}, S.~F., {et~al.} 2013,
  \aap, 559, A114

\bibitem[{{McMullin} {et~al.}(2007){McMullin}, {Waters}, {Schiebel}, {Young},
  \& {Golap}}]{McMullin2007}
{McMullin}, J.~P., {Waters}, B., {Schiebel}, D., {Young}, W., \& {Golap}, K.
  2007, in Astronomical Society of the Pacific Conference Series, Vol. 376,
  Astronomical Data Analysis Software and Systems XVI, ed. R.~A. {Shaw},
  F.~{Hill}, \& D.~J. {Bell}, 127

\bibitem[{{Micha{\l}owski} {et~al.}(2016){Micha{\l}owski}, {Castro Cer{\'o}n},
  {Wardlow}, {Karska}, {Messias}, {van der Werf}, {Hunt}, {Baes},
  {Castro-Tirado}, {Gentile}, {Hjorth}, {Le Floc'h},
  {P{\'e}rez-Mart{\'{\i}}nez}, {Nicuesa Guelbenzu}, {Rasmussen}, {Rizzo},
  {Rossi}, {S{\'a}nchez-Portal}, {Schady}, {Sollerman}, \&
  {Xu}}]{Michalowski2016}
{Micha{\l}owski}, M.~J., {Castro Cer{\'o}n}, J.~M., {Wardlow}, J.~L., {et~al.}
  2016, \aap, 595, A72

\bibitem[{{Micha{\l}owski} {et~al.}(2015){Micha{\l}owski}, {Gentile}, {Hjorth},
  {Krumholz}, {Tanvir}, {Kamphuis}, {Burlon}, {Baes}, {Basa}, {Berta}, {Castro
  Cer{\'o}n}, {Crosby}, {D'Elia}, {Elliott}, {Greiner}, {Hunt}, {Klose},
  {Koprowski}, {Le Floc'h}, {Malesani}, {Murphy}, {Nicuesa Guelbenzu},
  {Palazzi}, {Rasmussen}, {Rossi}, {Savaglio}, {Schady}, {Sollerman}, {de
  Ugarte Postigo}, {Watson}, {van der Werf}, {Vergani}, \&
  {Xu}}]{Michalowski2015}
{Micha{\l}owski}, M.~J., {Gentile}, G., {Hjorth}, J., {et~al.} 2015, \aap, 582,
  A78

\bibitem[{{Micha{\l}owski} {et~al.}(2012){Micha{\l}owski}, {Kamble}, {Hjorth},
  {Malesani}, {Reinfrank}, {Bonavera}, {Castro Cer{\'o}n}, {Ibar}, {Dunlop},
  {Fynbo}, {Garrett}, {Jakobsson}, {Kaplan}, {Kr{\"u}hler}, {Levan},
  {Massardi}, {Pal}, {Sollerman}, {Tanvir}, {van der Horst}, {Watson}, \&
  {Wiersema}}]{Michalowski2012}
{Micha{\l}owski}, M.~J., {Kamble}, A., {Hjorth}, J., {et~al.} 2012, \apj, 755,
  85

\bibitem[{{Micha{\l}owski} {et~al.}(2018{\natexlab{a}}){Micha{\l}owski},
  {Karska}, {Rizzo}, {Baes}, {Castro-Tirado}, {Hjorth}, {Hunt}, {Kamphuis},
  {Koprowski}, {Krumholz}, {Malesani}, {Nicuesa Guelbenzu}, {Rasmussen},
  {Rossi}, {Schady}, {Sollerman}, \& {van der Werf}}]{Michalowski2018}
{Micha{\l}owski}, M.~J., {Karska}, A., {Rizzo}, J.~R., {et~al.}
  2018{\natexlab{a}}, \aap, 617, A143

\bibitem[{{Micha{\l}owski} {et~al.}(2018{\natexlab{b}}){Micha{\l}owski}, {Xu},
  {Stevens}, {Levan}, {Yang}, {Paragi}, {Kamble}, {Tsai}, {Dannerbauer}, {van
  der Horst}, {Shao}, {Crosby}, {Gentile}, {Stanway}, {Wiersema}, {Fynbo},
  {Tanvir}, {Kamphuis}, {Garrett}, \& {Bartczak}}]{Michalowski18grb}
{Micha{\l}owski}, M.~J., {Xu}, D., {Stevens}, J., {et~al.} 2018{\natexlab{b}},
  \aap, 616, A169

\bibitem[{{Mirzoyan} {et~al.}(2019){Mirzoyan}, {Noda}, {Moretti}, {Berti},
  {Nigro}, {Hoang}, {Micanovic}, {Takahashi}, {Chai}, \&
  {Moralejo}}]{Mirzoyan2019}
{Mirzoyan}, R., {Noda}, K., {Moretti}, E., {et~al.} 2019, GRB Coordinates
  Network, 23701

\bibitem[{{Modjaz} {et~al.}(2011){Modjaz}, {Kewley}, {Bloom}, {Filippenko},
  {Perley}, \& {Silverman}}]{Modjaz2011}
{Modjaz}, M., {Kewley}, L., {Bloom}, J.~S., {et~al.} 2011, \apjl, 731, L4

\bibitem[{{Nava}(2018)}]{Nava2018}
{Nava}, L. 2018, International Journal of Modern Physics D, 27, 1842003

\bibitem[{{Osterbrock}(1989)}]{Osterbrock1989}
{Osterbrock}, D.~E. 1989, {Astrophysics of gaseous nebulae and active galactic
  nuclei}

\bibitem[{{Palmerio} {et~al.}(2019){Palmerio}, {Vergani}, {Salvaterra}, {Sand
  ers}, {Japelj}, {Vidal-Garc{\'\i}a}, {D'Avanzo}, {Corre}, {Perley},
  {Shapley}, {Boissier}, {Greiner}, {Le Floc'h}, \& {Wiseman}}]{Palmerio2019}
{Palmerio}, J.~T., {Vergani}, S.~D., {Salvaterra}, R., {et~al.} 2019, \aap,
  623, A26

\bibitem[{{Pe'er} \& {Waxman}(2005)}]{Peer2005}
{Pe'er}, A. \& {Waxman}, E. 2005, \apj, 633, 1018

\bibitem[{{Pei}(1992)}]{Pei1992}
{Pei}, Y.~C. 1992, \apj, 395, 130

\bibitem[{{Perley} {et~al.}(2016{\natexlab{a}}){Perley}, {Kr{\"u}hler},
  {Schulze}, {de Ugarte Postigo}, {Hjorth}, {Berger}, {Cenko}, {Chary},
  {Cucchiara}, \& {Ellis}}]{Perley2016A}
{Perley}, D.~A., {Kr{\"u}hler}, T., {Schulze}, S., {et~al.} 2016{\natexlab{a}},
  \apj, 817, 7

\bibitem[{{Perley} {et~al.}(2016{\natexlab{b}}){Perley}, {Tanvir}, {Hjorth},
  {Laskar}, {Berger}, {Chary}, {de Ugarte Postigo}, {Fynbo}, {Kr{\"u}hler}, \&
  {Levan}}]{Perley2016B}
{Perley}, D.~A., {Tanvir}, N.~R., {Hjorth}, J., {et~al.} 2016{\natexlab{b}},
  \apj, 817, 8

\bibitem[{{Pirard} {et~al.}(2004){Pirard}, {Kissler-Patig}, {Moorwood},
  {Biereichel}, {Delabre}, {Dorn}, {Finger}, {Gojak}, {Huster}, {Jung}, {Koch},
  {Le Louarn}, {Lizon}, {Mehrgan}, {Pozna}, {Silber}, {Sokar}, \&
  {Stegmeier}}]{pirard2004}
{Pirard}, J.-F., {Kissler-Patig}, M., {Moorwood}, A., {et~al.} 2004, in
  \procspie, Vol. 5492, Ground-based Instrumentation for Astronomy, ed.
  A.~F.~M. {Moorwood} \& M.~{Iye}, 1763--1772

\bibitem[{{S{\'a}nchez-Ram{\'\i}rez} {et~al.}(2017){S{\'a}nchez-Ram{\'\i}rez},
  {Hancock}, {J{\'o}hannesson}, {Murphy}, {de Ugarte Postigo}, {Gorosabel},
  {Kann}, {Kr{\"u}hler}, {Oates}, {Japelj}, {Th{\"o}ne}, {Lundgren}, {Perley},
  {Malesani}, {de Gregorio Monsalvo}, {Castro-Tirado}, {D'Elia}, {Fynbo},
  {Garcia-Appadoo}, {Goldoni}, {Greiner}, {Hu}, {Jel{\'\i}nek}, {Jeong},
  {Kamble}, {Klose}, {Kuin}, {Llorente}, {Mart{\'\i}n}, {Nicuesa Guelbenzu},
  {Rossi}, {Schady}, {Sparre}, {Sudilovsky}, {Tello}, {Updike}, {Wiersema}, \&
  {Zhang}}]{SanchezRamirez2017}
{S{\'a}nchez-Ram{\'\i}rez}, R., {Hancock}, P.~J., {J{\'o}hannesson}, G.,
  {et~al.} 2017, \mnras, 464, 4624

\bibitem[{{Sanders} {et~al.}(1991){Sanders}, {Scoville}, \&
  {Soifer}}]{Sanders1991}
{Sanders}, D.~B., {Scoville}, N.~Z., \& {Soifer}, B.~T. 1991, \apj, 370, 158

\bibitem[{{Sanders} {et~al.}(2012){Sanders}, {Soderberg}, {Levesque}, {Foley},
  {Chornock}, {Milisavljevic}, {Margutti}, {Berger}, {Drout}, {Czekala}, \&
  {Dittmann}}]{Sanders2012}
{Sanders}, N.~E., {Soderberg}, A.~M., {Levesque}, E.~M., {et~al.} 2012, \apj,
  758, 132

\bibitem[{{Schady} {et~al.}(2015){Schady}, {Kr{\"u}hler}, {Greiner}, {Graham},
  {Kann}, {Bolmer}, {Delvaux}, {Elliott}, {Klose}, {Knust}, {Nicuesa
  Guelbenzu}, {Rau}, {Rossi}, {Savaglio}, {Schmidl}, {Schweyer}, {Sudilovsky},
  {Tanga}, {Tanvir}, {Varela}, \& {Wiseman}}]{Schady2015}
{Schady}, P., {Kr{\"u}hler}, T., {Greiner}, J., {et~al.} 2015, \aap, 579, A126

\bibitem[{{Schady} {et~al.}(2014){Schady}, {Savaglio}, {M{\"u}ller},
  {Kr{\"u}hler}, {Dwelly}, {Palazzi}, {Hunt}, {Greiner}, {Linz},
  {Micha{\l}owski}, {Pierini}, {Piranomonte}, {Vergani}, \&
  {Gear}}]{Schady2014}
{Schady}, P., {Savaglio}, S., {M{\"u}ller}, T., {et~al.} 2014, \aap, 570, A52

\bibitem[{Schlafly \& Finkbeiner(2011)}]{Schlafly2011}
Schlafly, E.~F. \& Finkbeiner, D.~P. 2011, ApJ, 737, 103

\bibitem[{{Scudder} {et~al.}(2012){Scudder}, {Ellison}, {Torrey}, {Patton}, \&
  {Mendel}}]{Scudder2012}
{Scudder}, J.~M., {Ellison}, S.~L., {Torrey}, P., {Patton}, D.~R., \& {Mendel},
  J.~T. 2012, \mnras, 426, 549

\bibitem[{{Selsing} {et~al.}(2019){Selsing}, {Fynbo}, {Heintz}, {Watson}, \&
  {Dyrbye}}]{Selsing2019}
{Selsing}, J., {Fynbo}, J.~P.~U., {Heintz}, K.~E., {Watson}, D., \& {Dyrbye},
  N. 2019, GRB Coordinates Network, 23695

\bibitem[{{Sirianni} {et~al.}(2005){Sirianni}, {Jee}, {Ben{\'\i}tez},
  {Blakeslee}, {Martel}, {Meurer}, {Clampin}, {De Marchi}, {Ford}, \&
  {Gilliland}}]{Sirianni2005}
{Sirianni}, M., {Jee}, M.~J., {Ben{\'\i}tez}, N., {et~al.} 2005, \pasp, 117,
  1049

\bibitem[{{Skrutskie} {et~al.}(2006){Skrutskie}, {Cutri}, {Stiening},
  {Weinberg}, {Schneider}, {Carpenter}, {Beichman}, {Capps}, {Chester},
  {Elias}, {Huchra}, {Liebert}, {Lonsdale}, {Monet}, {Price}, {Seitzer},
  {Jarrett}, {Kirkpatrick}, {Gizis}, {Howard}, {Evans}, {Fowler}, {Fullmer},
  {Hurt}, {Light}, {Kopan}, {Marsh}, {McCallon}, {Tam}, {Van Dyk}, \&
  {Wheelock}}]{Skrutskie2006AJ}
{Skrutskie}, M.~F., {Cutri}, R.~M., {Stiening}, R., {et~al.} 2006, \aj, 131,
  1163

\bibitem[{{Solomon} {et~al.}(1997){Solomon}, {Downes}, {Radford}, \&
  {Barrett}}]{Solomon1997}
{Solomon}, P.~M., {Downes}, D., {Radford}, S.~J.~E., \& {Barrett}, J.~W. 1997,
  \apj, 478, 144

\bibitem[{{Spergel} {et~al.}(2003){Spergel}, {Verde}, {Peiris}, {Komatsu},
  {Nolta}, {Bennett}, {Halpern}, {Hinshaw}, {Jarosik}, {Kogut}, {Limon},
  {Meyer}, {Page}, {Tucker}, {Weiland}, {Wollack}, \& {Wright}}]{Spergel2003}
{Spergel}, D.~N., {Verde}, L., {Peiris}, H.~V., {et~al.} 2003, \apjs, 148, 175

\bibitem[{{Stanway} {et~al.}(2015){Stanway}, {Levan}, {Tanvir}, {Wiersema}, \&
  {van der Laan}}]{Stanway2015}
{Stanway}, E.~R., {Levan}, A.~J., {Tanvir}, N.~R., {Wiersema}, K., \& {van der
  Laan}, T.~P.~R. 2015, \apjl, 798, L7

\bibitem[{{Svensson} {et~al.}(2010){Svensson}, {Levan}, {Tanvir}, {Fruchter},
  \& {Strolger}}]{Svensson2010}
{Svensson}, K.~M., {Levan}, A.~J., {Tanvir}, N.~R., {Fruchter}, A.~S., \&
  {Strolger}, L.~G. 2010, \mnras, 405, 57

\bibitem[{{Takahashi}(2019)}]{Takahashi2019ICRC}
{Takahashi}, M. 2019, in International Cosmic Ray Conference, Vol.~36, 36th
  International Cosmic Ray Conference (ICRC2019), 606

\bibitem[{{Tanvir} {et~al.}(2004){Tanvir}, {Barnard}, {Blain}, {Fruchter},
  {Kouveliotou}, {Natarajan}, {Ramirez-Ruiz}, {Rol}, {Smith}, {Tilanus}, \&
  {Wijers}}]{Tanvir2004}
{Tanvir}, N.~R., {Barnard}, V.~E., {Blain}, A.~W., {et~al.} 2004, \mnras, 352,
  1073

\bibitem[{{The Fermi GBM team}(2019)}]{Fermi2019GCN}
{The Fermi GBM team}. 2019, GRB Coordinates Network, 25551, 1

\bibitem[{{Tokunaga} \& {Vacca}(2005)}]{Tokunaga2005}
{Tokunaga}, A.~T. \& {Vacca}, W.~D. 2005, \pasp, 117, 421

\bibitem[{{Vergani} {et~al.}(2015){Vergani}, {Salvaterra}, {Japelj}, {Le
  Floc'h}, {D'Avanzo}, {Fernandez-Soto}, {Kr{\"u}hler}, {Melandri}, {Boissier},
  {Covino}, {Puech}, {Greiner}, {Hunt}, {Perley}, {Petitjean}, {Vinci},
  {Hammer}, {Levan}, {Mannucci}, {Campana}, {Flores}, {Gomboc}, \&
  {Tagliaferri}}]{Vergani2015}
{Vergani}, S.~D., {Salvaterra}, R., {Japelj}, J., {et~al.} 2015, \aap, 581,
  A102

\bibitem[{{Vernet} {et~al.}(2011){Vernet}, {Dekker}, {D'Odorico}, {Kaper},
  {Kjaergaard}, {Hammer}, {Randich}, {Zerbi}, {Groot}, {Hjorth}, {Guinouard},
  {Navarro}, {Adolfse}, {Albers}, {Amans}, {Andersen}, {Andersen}, {Binetruy},
  {Bristow}, {Castillo}, {Chemla}, {Christensen}, {Conconi}, {Conzelmann},
  {Dam}, {de Caprio}, {de Ugarte Postigo}, {Delabre}, {di Marcantonio},
  {Downing}, {Elswijk}, {Finger}, {Fischer}, {Flores}, {Fran{\c c}ois},
  {Goldoni}, {Guglielmi}, {Haigron}, {Hanenburg}, {Hendriks}, {Horrobin},
  {Horville}, {Jessen}, {Kerber}, {Kern}, {Kiekebusch}, {Kleszcz}, {Klougart},
  {Kragt}, {Larsen}, {Lizon}, {Lucuix}, {Mainieri}, {Manuputy}, {Martayan},
  {Mason}, {Mazzoleni}, {Michaelsen}, {Modigliani}, {Moehler}, {M{\o}ller},
  {Norup S{\o}rensen}, {N{\o}rregaard}, {P{\'e}roux}, {Patat}, {Pena}, {Pragt},
  {Reinero}, {Rigal}, {Riva}, {Roelfsema}, {Royer}, {Sacco}, {Santin},
  {Schoenmaker}, {Spano}, {Sweers}, {Ter Horst}, {Tintori}, {Tromp}, {van
  Dael}, {van der Vliet}, {Venema}, {Vidali}, {Vinther}, {Vola}, {Winters},
  {Wistisen}, {Wulterkens}, \& {Zacchei}}]{Vernet2011}
{Vernet}, J., {Dekker}, H., {D'Odorico}, S., {et~al.} 2011, \aap, 536, A105

\bibitem[{{Violino} {et~al.}(2018){Violino}, {Ellison}, {Sargent}, {Coppin},
  {Scudder}, {Mendel}, \& {Saintonge}}]{Violino2018}
{Violino}, G., {Ellison}, S.~L., {Sargent}, M., {et~al.} 2018, \mnras, 476,
  2591

\bibitem[{{Wang} {et~al.}(2012){Wang}, {Chen}, \& {Huang}}]{Wang2012}
{Wang}, W.-H., {Chen}, H.-W., \& {Huang}, K.-Y. 2012, \apjl, 761, L32

\bibitem[{{Wang} {et~al.}(2010){Wang}, {He}, {Li}, {Wu}, \& {Dai}}]{Wang2010}
{Wang}, X.-Y., {He}, H.-N., {Li}, Z., {Wu}, X.-F., \& {Dai}, Z.-G. 2010, \apj,
  712, 1232

\bibitem[{{Whitaker} {et~al.}(2014){Whitaker}, {Franx}, {Leja}, {van Dokkum},
  {Henry}, {Skelton}, {Fumagalli}, {Momcheva}, {Brammer}, {Labb{\'e}},
  {Nelson}, \& {Rigby}}]{Whitaker14}
{Whitaker}, K.~E., {Franx}, M., {Leja}, J., {et~al.} 2014, \apj, 795, 104

\bibitem[{{Wright} {et~al.}(2010){Wright}, {Eisenhardt}, {Mainzer}, {Ressler},
  {Cutri}, {Jarrett}, {Kirkpatrick}, {Padgett}, {McMillan}, {Skrutskie},
  {Stanford}, {Cohen}, {Walker}, {Mather}, {Leisawitz}, {Gautier}, {McLean},
  {Benford}, {Lonsdale}, {Blain}, {Mendez}, {Irace}, {Duval}, {Liu}, {Royer},
  {Heinrichsen}, {Howard}, {Shannon}, {Kendall}, {Walsh}, {Larsen}, {Cardon},
  {Schick}, {Schwalm}, {Abid}, {Fabinsky}, {Naes}, \& {Tsai}}]{Wright2010}
{Wright}, E.~L., {Eisenhardt}, P.~R.~M., {Mainzer}, A.~K., {et~al.} 2010, \aj,
  140, 1868

\bibitem[{{Wuyts} {et~al.}(2014){Wuyts}, {Kurk}, {F{\"o}rster Schreiber},
  {Genzel}, {Wisnioski}, {Bandara}, {Wuyts}, {Beifiori}, {Bender}, {Brammer},
  {Burkert}, {Buschkamp}, {Carollo}, {Chan}, {Davies}, {Eisenhauer}, {Fossati},
  {Kulkarni}, {Lang}, {Lilly}, {Lutz}, {Mancini}, {Mendel}, {Momcheva}, {Naab},
  {Nelson}, {Renzini}, {Rosario}, {Saglia}, {Seitz}, {Sharples}, {Sternberg},
  {Tacchella}, {Tacconi}, {van Dokkum}, \& {Wilman}}]{Wuyts14}
{Wuyts}, E., {Kurk}, J., {F{\"o}rster Schreiber}, N.~M., {et~al.} 2014, \apjl,
  789, L40

\bibitem[{{Wuyts} {et~al.}(2016){Wuyts}, {Wisnioski}, {Fossati}, {F{\"o}rster
  Schreiber}, {Genzel}, {Davies}, {Mendel}, {Naab}, {R{\"o}ttgers}, {Wilman},
  {Wuyts}, {Bandara}, {Beifiori}, {Belli}, {Bender}, {Brammer}, {Burkert},
  {Chan}, {Galametz}, {Kulkarni}, {Lang}, {Lutz}, {Momcheva}, {Nelson},
  {Rosario}, {Saglia}, {Seitz}, {Tacconi}, {Tadaki}, {{\"U}bler}, \& {van
  Dokkum}}]{Wuyts16}
{Wuyts}, E., {Wisnioski}, E., {Fossati}, M., {et~al.} 2016, \apj, 827, 74

\bibitem[{{Young} {et~al.}(1989){Young}, {Xie}, {Kenney}, \&
  {Rice}}]{Young1989}
{Young}, J.~S., {Xie}, S., {Kenney}, J. D.~P., \& {Rice}, W.~L. 1989, \apjs,
  70, 699

\bibitem[{{Zahid} {et~al.}(2013){Zahid}, {Yates}, {Kewley}, \&
  {Kudritzki}}]{Zahid13}
{Zahid}, H.~J., {Yates}, R.~M., {Kewley}, L.~J., \& {Kudritzki}, R.~P. 2013,
  \apj, 763, 92

\end{thebibliography}

\end{document}